\documentclass[twocolumn]{IEEEtran}
\normalsize
\usepackage{amsmath, amssymb}
\usepackage{amsthm}
\usepackage{mathtools}
\usepackage[small]{caption}
\usepackage{amsthm,multirow,color,amsfonts}
\usepackage[ruled,linesnumbered]{algorithm2e}
\usepackage{tabulary}
\usepackage{subfigure}
\usepackage{graphicx}
\usepackage{setspace}
\usepackage{enumerate}
\usepackage{comment}
\usepackage{multicol,lipsum}
\usepackage{cuted}
\usepackage{cite}
\usepackage{bm}
\usepackage{bbm}
\usepackage[noend]{algpseudocode}
\usepackage{hyperref}
\makeatletter
\def\BState{\State\hskip-\ALG@thistlm}
\makeatother
	
\newtheorem{theo}{Theorem}
\newtheorem{defi}{Definition}

\newtheorem{prop}{Proposition}

\newtheorem{lem}{Lemma}

\usepackage{soul}

\IEEEaftertitletext{\vspace{-1\baselineskip}}

\title{Delay-Optimal Forwarding and Computation Offloading for Service Chain Tasks}
\author{Jinkun Zhang, Yuezhou Liu, and Edmund Yeh
\thanks{The authors are with Electrical and Computer Engineering Department, Northeastern University, Boston, USA (e-mail:zhang.jinku@northeastern.edu, liu.yuez@northeastern.edu, eyeh@ece.neu.edu).} 
}

\IEEEoverridecommandlockouts
\begin{document}
\maketitle
\begin{abstract}
Emerging edge computing paradigms enable heterogeneous devices to collaborate on complex computation applications.
However, for congestible links and computing units, delay-optimal forwarding and offloading for service chain tasks (e.g., DNN with vertical split) in edge computing networks remains an open problem.
In this paper, we formulate the service chain forwarding and offloading problem with arbitrary topology and heterogeneous transmission/computation capability, and aim to minimize the aggregated network cost. 
We consider congestion-aware nonlinear cost functions that cover various performance metrics and constraints, such as average queueing delay with limited processor capacity.
We solve the non-convex optimization problem globally by analyzing the KKT condition and proposing a sufficient condition for optimality.
We then propose a distributed algorithm that converges to the global optimum. The algorithm adapts to changes in input rates and network topology, and can be implemented as an online algorithm.
Numerical evaluation shows that our method significantly outperforms baselines in multiple network instances, especially in congested scenarios.
\end{abstract}


\section{Introduction}
\label{Section:introduction}
Recent years have seen an explosion in the number of mobile and IoT devices. Many of the emerging mobile applications, such as VR/AR, autonomous driving, are computation-intensive and time-critical.
Mobile devices running these applications generate a huge amount of data traffic, which is predicted to reach 288EB per month in 2027 \cite{Ericsson2021report}. 
It is becoming impractical to direct all computation requests and their data to the central cloud due to limited backhaul bandwidth and high associated latency.
Edge computing has been proposed as a promising solution to provide computation resources and cloud-like services in close proximity to mobile devices.   
Well-known edge computing paradigms include mobile edge computing (MEC) and fog computing, which deploy computation resources at wireless access points and gateways, respectively. 

In edge computing, requesters offload their computation to the edge servers, where the network topology is typically hierarchical. 
Extending the idea of edge computing is a new concept called collaborative edge computing (CEC), in which the network structure is more flexible.
In addition to point-to-point offloading, CEC permits multiple stakeholders (mobile devices, IoT devices, edge servers, or cloud) to collaborate with each other by sharing data, communication resources, and computation resources to accomplish computation tasks \cite{sahni2020multi}. CEC improves the utilization efficiency of resources so that computation-intensive and time-critical services can be better completed at the edge. Mobile devices equipped with computation capabilities can collaborate with each other through D2D communication \cite{sahni2017edge}. Edge servers can also collaborate with each other 
for load balancing or further with the central cloud to offload demands that they cannot accommodate \cite{zhu2017socially}. Furthermore, CEC is needed when there is no direct connection between devices and edge servers. Consider unmanned aerial vehicle (UAV) swarms or autonomous cars in rural areas, computation-intensive tasks of UAVs or cars far away from the wireless access point should be collaboratively computed or offloaded through multi-hop routing to the server with the help of other devices \cite{hong2019multi,sahni2017edge}. 

{
The global optimal forwarding and task offloading strategy in CEC has been proposed \cite{zhang2022optimal}, however, restricted to single-step computation tasks.
With the development of network function virtualization (NFV), \emph{Service chain} is a generalization of the traditional single-step computation, enabling the ordering of computation tasks \cite{han2015network}.
In the service chain model, the network provides services of \emph{applications}, where each application consists of a chain of \emph{tasks} that are performed sequentially, mapping input data to output results with potentially multiple intermediate stages.
Fig \ref{fig_service_chain} exemplifies service chaining with a video stream client for a local area network (LAN).
Input streaming data may undergo two tasks before reaching display: a firewall and a transcoder. 
The firewall accepts only authorized data; the transcoder decompresses the video and adjusts the resolution as user-specified.
With CEC, such service chain can be accomplished in a distributed manner, where the tasks are collaboratively offloaded in the LAN without requiring a centralized video-processing server. 

\begin{figure}[htbp]
\centerline{\includegraphics[width=0.4\textwidth]{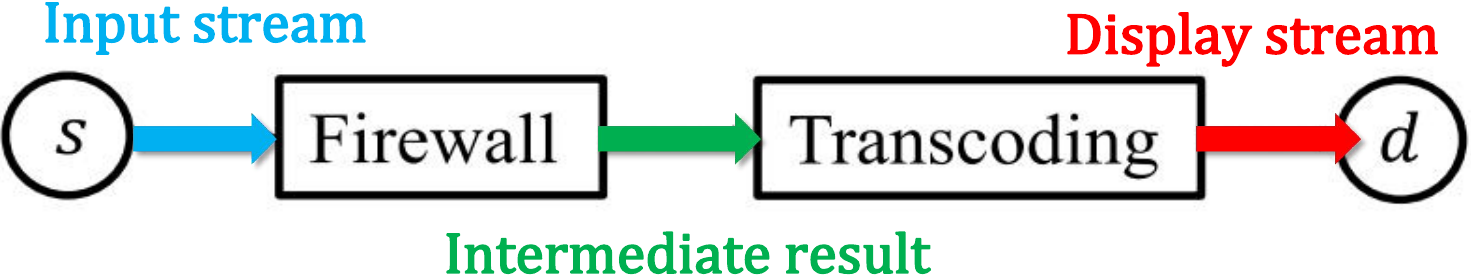}}
\caption{Service chain example: LAN video stream client from source $s$ to display $d$.  Input data stream undergoes two sequential tasks.}
\label{fig_service_chain}
\end{figure}

The service chain model has been widely studied in the NFV context, e.g., \cite{zhang2021optimal}, but has not yet been introduced to CEC.
}
In this paper, we study a general framework of CEC that enables various types of collaboration for service chain applications among stakeholders. 
We consider a multi-hop network with arbitrary topology, where nodes collaboratively finish multiple {service chain applications}. 
Nodes have heterogeneous computation capabilities, and some are also data sources (e.g., sensors and mobile users) that generate data for computation.
{For each application, data are input from potentially multiple sources, intermediate results are acquired and forwarded among all collaborative nodes to achieve a minimum network resource usage, and the final result is delivered to the requester node. }
We allow partial offloading introduced in \cite{wang2016mobile}, i.e., a task can be partitioned into multiple components and separately offloaded. 
For example, in video compression, the original video can be chunked into blocks and compressed separately at multiple devices, and the results could then be merged.

{
Performing an application requires 
1) forwarding data from the sources to multiple nodes for computation; 
2) forwarding intermediate results to potentially different nodes for next-stage computation; 
3) forwarding final results to the destination. 
We aim for a joint forwarding (how to route data and results) and computation offloading (where to perform computation tasks of the service chain) strategy that minimizes the total communication and computation costs.
}

{
We propose a framework that unifies data/result forwarding and computation placement, and formulate the joint forwarding and offloading problem for service chain applications as a non-convex optimization problem.
We tackle the problem with a node-based perspective first introduced by \cite{gallager1977minimum} and followed by \cite{xi2008node}\cite{zhang2022optimal}.
We first investigate the Karush–Kuhn–Tucker (KKT) necessary conditions for the proposed optimization problem, and demonstrate by example that such KKT condition can lead to { arbitrarily suboptimal performance}.
To address this, we then devise a set of sufficient conditions for global optimality. 
To obtain the sufficient condition, we adopt a modification technique to the KKT condition first invented by Gallager et al. \cite{gallager1977minimum} and followed by many subsequent works \cite{mahdian2018mindelay, zhang2023congestion, zhang2022optimal}. 
To our knowledge, however, no previous work has provided a theoretical foundation for this modification technique that guarantees a global solution for a non-convex problem.
We fill this gap by showing that our non-convex objective is geodesically convex under the assumption that input rates are strictly positive. 
We are the first to reveal such mathematical structure in network optimization problems.
We demonstrate that the geodesic convexity holds for a class of generalized problems, including joint optimization with fairness and congestion control.

Based on the sufficient optimality condition, 
we propose a node-based algorithm that supports distributed and online implementation, where nodes need only exchange information with their immediate neighbors.
The algorithm is adaptive to moderate changes in network parameters, including input rates, performance metrics, etc.
}

Our detailed contributions are:
\begin{itemize}
    \item We formulate forwarding and computation offloading for service chain applications with arbitrary network topology and congestible links by a non-convex cost minimization problem.
    We show that the proposed problem is geodesically convex under additional assumptions.
    \item We propose a set of sufficient optimality conditions for the proposed non-convex problem, and devise a distributed and adaptive algorithm that converges to the proposed sufficient condition.
    {\item We extend our results to consider network congestion control and fairness jointly.}
    \item Through numerical experimentation, we show the advantages of the proposed algorithm over baselines in different network instances, especially in congested scenarios.
\end{itemize}

This paper is organized as follows:
In Section \ref{Section:relatedwork} we briefly compare this paper with related works.
To better comprehend our model, we first study the delay optimal forwarding and offloading for single-step computations in Section \ref{sec:single step}.
We then extend our model and technical results to general service chain applications in Section \ref{sec:service chain}.
Based on our analytical result, in Section \ref{Section:algorithm}, we propose a distributed and adaptive algorithm that converges to the global optimal solution.
Our numerical experimentation results are provided in Section \ref{Section:simulation}.
Finally, we discuss extensions of our method to incorporate congestion control and fairness in Section \ref{Section:extension}.


\begin{figure}[t]
\centerline{\includegraphics[width=0.45\textwidth]{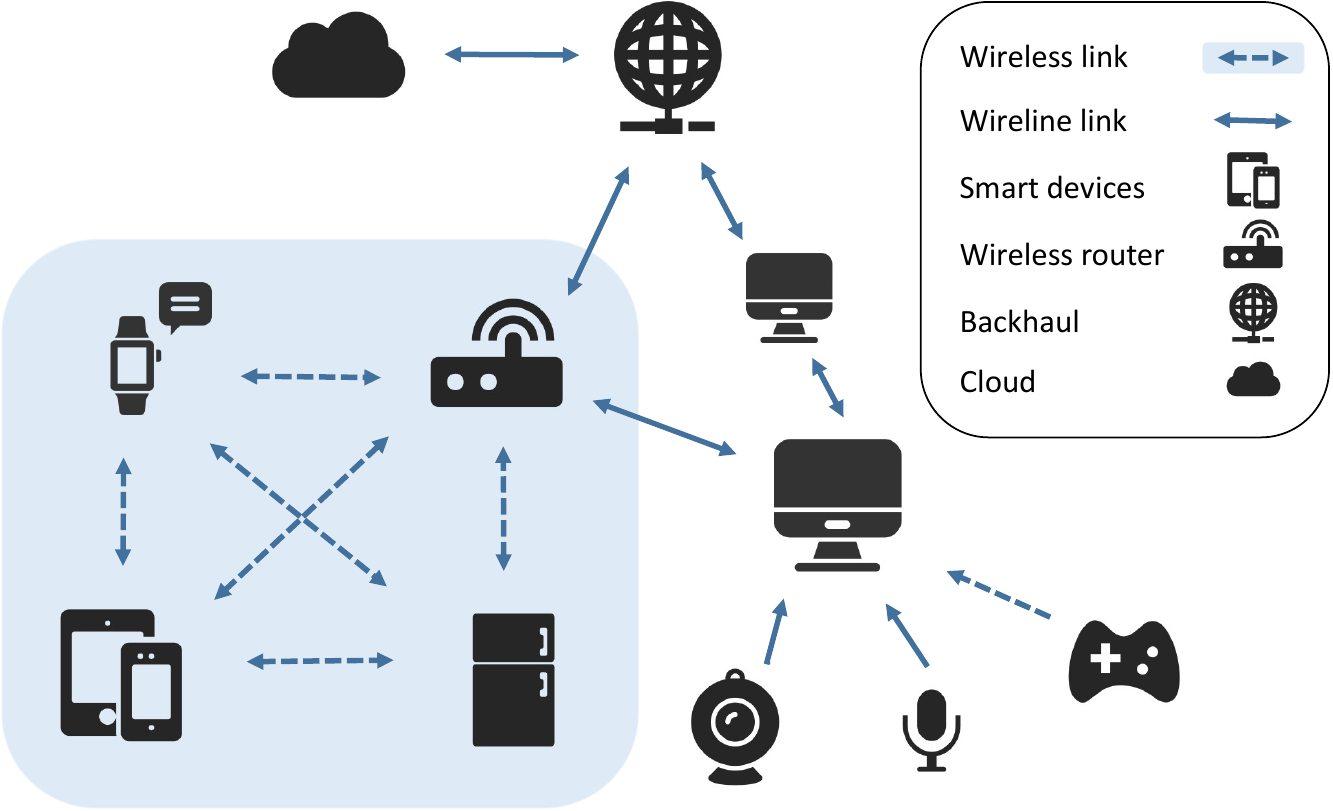}}
\caption{Sample system topology involving IoT network on the edge}
\label{fig1}
\end{figure}

\section{Related Works}
\label{Section:relatedwork}
\noindent\textbf{Computation offloading in edge computing.}
{
Joint routing and computation offloading in edge computing have been investigated in a number of prior contributions.  
Sahni et al. \cite{sahni2020multi} \cite{sahni2018data} adopt the model where each task is only performed once with the exact release time known, which we refer as \emph{single-instance model}. 
Zhang et al. \cite{zhang2021optimal} consider a \emph{flow model} where data collection and computation of each task are performed continuously, and time-averaged costs are measured based on the rates of data and result flows. 
Most existing studies in CEC assume the requester itself provides the data of tasks \cite{hong2019multi}\cite{liu2020distributed}\cite{al2016distributed}. 
Although Sahni et al. \cite{sahni2018data}\cite{sahni2017edge} %
consider arbitrary data sources, the network is assumed to be fully connected, or with predefined routing paths.
The communication cost in previous studies is often assumed to be solely due to input data transmission \cite{luo2021qoe}\cite{sahni2020multi}, and the results are transmitted simply along the reverse path of input data 
\cite{he2021multi}\cite{funai2019computational}.
However, the size of computation results is not negligible in many practical applications,
e.g., federated or distributed machine learning~\cite{mcmahan2017communication}, file decompression, and image enhancement.
For communication, Hong et al. \cite{hong2019multi,hong2019qos} formulate heterogeneous link transmission speeds, and Sahni et al. \cite{sahni2018data} consider link bandwidth constraints. However, most works assume the link costs be linear functions of the traffic.
{Xiang et al.~\cite{xiang2020joint} study routing and computation offloading jointly with network slicing, and propose a heuristic algorithm. They consider a flow model, with non-linear delay functions, but without considering computation results. }
{
Kamran et al. \cite{kamran2019deco} generalized the throughput-optimal ICN caching-forwarding mechanism \cite{yeh2014vip} to computation placement in information-centric networking (ICN).
The throughput optimality is preserved for arbitrary network topology by adopting the idea of backpressure \cite{tassiulas1990stability}. 
}

Distinct from the above studies, in this paper, our formulation simultaneously 1) adopts the flow model on edge computing networks with arbitrary multi-hop topology, 2) allows the requester node to be distinct from data sources, 
3) optimizes forwarding for both data and (intermediate) results of non-negligible size,
and 4) models network congestion by considering non-linear communication and computation costs.

\vspace{0.5\baselineskip}
\noindent\textbf{Optimization of service chain applications.}
{
Service function chaining (SFC) was introduced by Halpern et al. \cite{halpern2015service} as a network architecture.
It is widely used in the NFV context to capture the dependence among network functions, and is followed with implementations, e.g., by dynamic backpressure by Kulkarni et al. \cite{kulkarni2017nfvnice}.
Qu et al. \cite{qu2016delay} studied NFV scheduling and resource allocation that optimize for the earliest finish time for service chain applications. 
Zhang et al. \cite{zhang2021optimal} proposed a throughput-optimal control policy for distributed computing networks with service chain applications and mixed-cast flows.
Recently with emerging learning techniques, service chain computation offloading is further developed beyond traditional optimization techniques.
Khoramnejad et al. \cite{khoramnejad2021distributed} minimized the cost of service chain delay and energy consumption in the edge via deep learning, but restricted to bipartite networks.
Wang et al. \cite{wang2020adaptive} propose a task chain scheduling algorithm for SFC allocation in the NFV-enabled MEC system via deep reinforcement learning.
This paper is distinct from the above studies as we are the first to directly minimize time-averaged user latency with queueing effect for service chains.

It is worth noting that, the service chain application model can be further generalized to capture even more complex interdependency among tasks, e.g., a directed acyclic graph (DAG) \cite{yang2019communication} enables the molding of reusable data and multi-input computation jobs.
With more complex interdependency, coding techniques can be applied to further exploit data/result reuse and reduce redundancy, where the theoretical bound is provided by Li et al. \cite{li2017coding}, and the practical implementation is built by Ghosh et al. \cite{ghosh2021jupiter}.
However, this paper focuses on the service chain model, of which our current methodology guarantees global optimality. Generalizing to DAG computation is considered as a future extension.
}

}

\section{Offloading of single-step computations}
\label{sec:single step}
\begin{figure}[htbp]
\centerline{\includegraphics[width=0.27\textwidth]{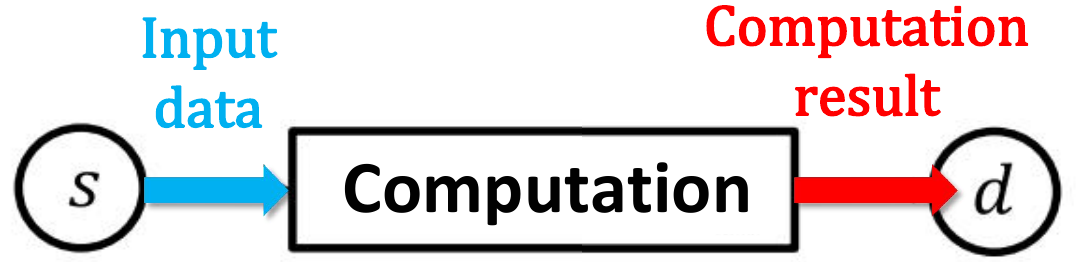}}
\caption{Single-step computation is a special case of service chain application with the chain length being $1$.}
\label{fig_singe_step}
\end{figure}

To help better understand our model and technical results, we start by presenting the delay-optimal forwarding and offloading for single-step computations.
As illustrated in Fig. \ref{fig_singe_step}, single-step computations can be viewed as a special case of service chain application where the chain length is $1$. 

We describe the network model and formally propose the optimization problem for single-step computations in Section \ref{Section:model_single}.
In Section \ref{sec:optimality_single}, we study the KKT necessity condition, and propose a sufficient optimality condition.

\subsection{Singe-step Computation Network}
\label{Section:model_single}
We begin by presenting our formal model of a collaborative dispersed computing network where nodes collaborate to carry out single-step computation tasks. 
Such networks are motivated by real-word applications such as IoT networks, connected vehicles and UAV swarms.
An example that involves an IoT network at the edge is shown in Fig. \ref{fig1}. 

\vspace{0.5\baselineskip}
\noindent\textbf{Network and tasks.}
We model the network with a directed and strongly connected graph $\mathcal{G} = (\mathcal{V},\mathcal{E})$, where $\mathcal{V}$ is the set of nodes 
and $\mathcal{E}$ is the set of links. 
A node may generally represent any facility, server, or real/virtual device that the network operator wishes to treat as a whole.
For nodes $i,j \in \mathcal{V}$, we say link $(i,j) \in \mathcal{E}$ if there exists an available channel from $i$ to $j$ that can effectively transmit data and result packets.
We assume that links in $\mathcal{E}$ are bidirectional, i.e., for any $(i,j) \in \mathcal{E}$, it holds that $(j,i) \in \mathcal{E}$.
For simplicity, we do not model the noise and interference, and assume all channels are error-free.
Let $\mathcal{N}(i) = \left\{j \big| (j,i) \in \mathcal{E}\right\} = \left\{j \big| (i,j) \in \mathcal{E}\right\}$ be the neighbors of $i$.
Single-step computations are performed by the nodes, mapping input data to results of non-negligible size. 
Data and results for computation are transmitted through the links. 
Nodes and links are assumed to have heterogeneous computation and communication capabilities, respectively.

Communication and computation are task-driven, where a \emph{task} involves 1) forwarding input data from (potentially multiple) data sources to computation sites, 2) computing, and 3) delivering the results to a pre-specified destination. 
For example, in an IoT monitor application, data sources could be sensors on different smart devices and the destination could be a user's cellphone. The data collected from the sensors is analyzed and processed in the network before being delivered to the user. 
We assume there are $M \in \mathbb{N}^+$ different types of single-step computations are performed. 
We denote a task by a pair $(d,m)$,  where $d \in \mathcal{V}$ is the destination node and $m \in \{1,\cdots,M\}$ is the specified computation type. 
We denote the set of all tasks by $\mathcal{T}$.

To incorporate partial offloading \cite{wang2016mobile} and measure the time-averaged network performance, for computation type $m$, we assume the exogenous input data is chunked into \emph{data packets} of equal size $L_{m}^-$. 
Computation could be independently performed on each data packet, and a \emph{result packet} is generated accordingly. We assume for computation type $m$, the result packets are of equal size $L_{m}^+$.
Such assumption is adopted in many partial offloading studies that consider result size, e.g.\cite{he2021multi}, where typically $L_{m}^- \geq L_{m}^+$. 
We also allow $L_{m}^- \leq L_{m}^+$ if the result size is larger than input data, e.g., video rendering, image super-resolution or file decompression.

\vspace{0.5\baselineskip}
\noindent\textbf{Data and result flows.}  
In contrast to the single-instance model where each task is performed only once 
\cite{sahni2020multi}, we adopt a flow model similar to \cite{zhang2018optimal}.
We assume the exogenous input data packets of each task are continuously injected into the network in the form of flows with certain rates, and the computations are continuously performed.
In the network, flows of data packets, i.e. \emph{data flows}, are routed as computational input to nodes with computation resources. After computation, flows of result packets, i.e., \emph{result flows}, are generated and routed to corresponding destinations. 

We assume the exogenous input data packets of task $(d,m)$ are injected into the network with rate $r_i(d,m) \geq 0$ (packet/s) at node $i$.
Let $\boldsymbol{r} = [r_i(d,m)]_{i \in \mathcal{V}, (d,m)\in\mathcal{T}}$ be the vector of global input rates.
\footnote{Note that we allow multiple nodes $i$ for which $r_i(d,m) > 0$, representing multiple data sources; $r_d(d,m)$ could also be positive, representing computation offloading with locally provided data. } 
Let $f_{ij}^-(d,m) \geq 0$ denote the data flow rate (packet/s) on link $(i,j)$ for task $(d,m)$.
Let $g_i(d,m) \geq 0$ be the data flow rate (packet/s) forwarded to node $i$'s computation unit for task $(d,m)$, referred as the \emph{computational input rate}. 
Moreover, let $f_{ij}^+(d,m)$ be the result flow rate (packet/s) on $(i,j)$ for $(d,m)$.

\begin{figure}
\centerline{
\includegraphics[width=0.48\textwidth]{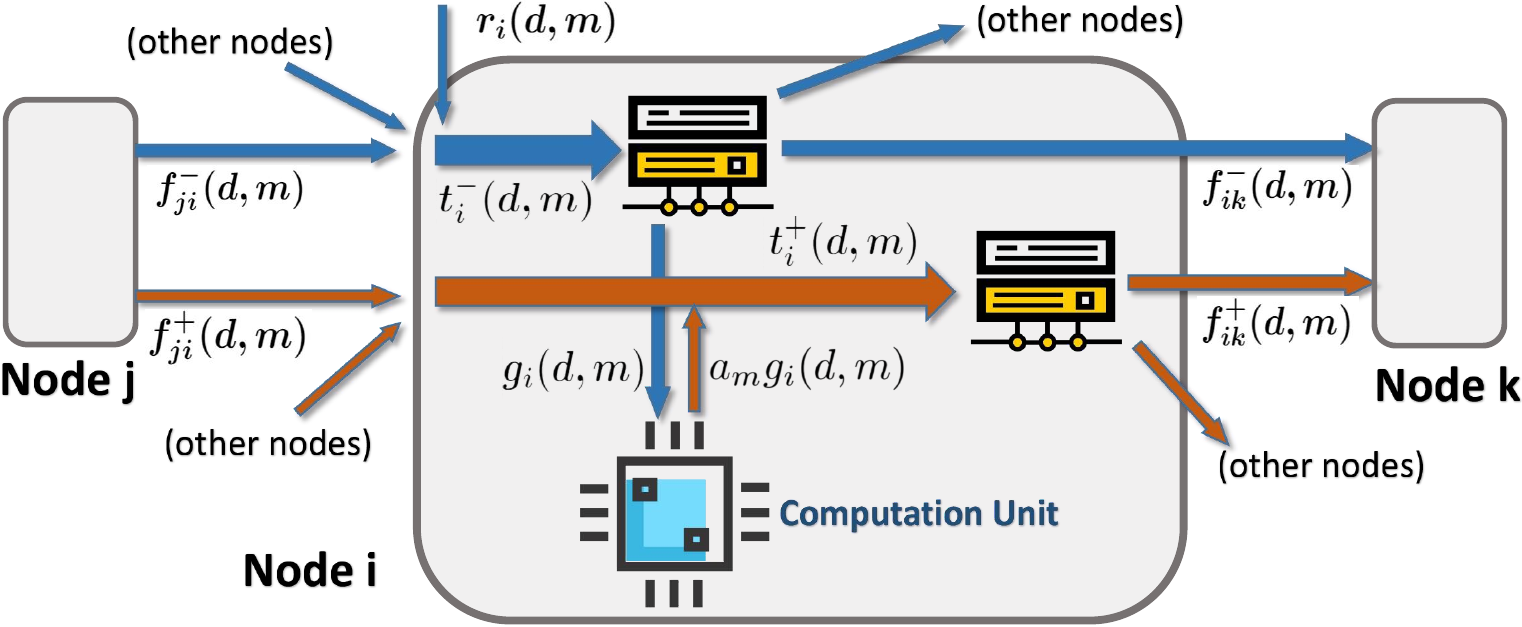}
}
\caption{Illustration of data and result forwarding for nodes $j \to i \to k$ for single-step computations }
\label{fig2}
\end{figure}

We let $t_i^-(d,m)$ and $t_i^+(d,m)$  be the total data rate and total result rate for task $(d,m)$ at node $i$, respectively,
\begin{align*}
      t_i^-(d,m) &= \sum\nolimits_{j \in \mathcal{N}(i)}f_{ji}^-(d,m) + r_i(d,m), 
    \\t_i^+(d,m) &= \sum\nolimits_{j \in \mathcal{N}(i)}f_{ji}^+(d,m) + g_i(d,m), 
\end{align*}
Fig.\ref{fig2}. illustrates our single-step computation flow model.

\vspace{0.5\baselineskip}
\noindent\textbf{Forwarding and offloading strategy.} 
The network performs hop-by-hop multi-path packet forwarding. 
For the forwarding of data flows, we let the forwarding variable $\phi_{ij}^-(d,m) \in [0,1] $ be the fraction of data packets forwarded to node $j$ by node $i$ for task $(d,m)$.
Namely, of the traffic $t_i^-(d,m)$, we assume node $i$ forwards a fraction of $\phi_{ij}^-(d,m) \in [0,1]$ to node $j \in \mathcal{N}(i)$.
Such fractional forwarding can be achieved via various methods, e.g., random packet dispatching, i.e., upon receiving a data packet for $d,m$, node $i$ randomly chooses one $j \mathcal{N}_i$ to forward, with probability $\phi_{ij}^-(d,m)$. 
More complex mechanisms also exist to stabilize the actual flow rates, e.g.,\cite{zhang2023congestion}.
Similarly, we let $\phi_{ij}^+(d,m) \in [0,1]$ be the forwarding variables of the result flow.
For computation offloading, let $\phi_{i0}^-(d,m)\in [0,1]$ be the fraction of data flow for task $(d,m)$ forwarded to the local computation unit of $i$.
Thus, 
\begin{equation*}
    \begin{aligned}
        f_{ij}^-(d,m) &= t_i^-(d,m) \phi_{ij}^-(d,m), \quad \forall j \in V 
    \\ g_i(d,m) &= t_i^-(d,m) \phi_{i0}^-(d,m), 
    \\ f_{ij}^+(d,m) &= t_i^+(d,m) \phi_{ij}^+(d,m). \quad\forall j \in V
    \end{aligned}
\end{equation*}
Note that $\phi_{ij}^{-}(d,m) = \phi_{ij}^{+}(d,m) \equiv 0$ if $(i,j) \not\in \mathcal{E}$. 
We denote by vector $\boldsymbol{\phi} = [\phi_{ij}^{-}(d,m), \phi_{ij}^{+}(d,m)]_{i,j \in \mathcal{V}, (d,m) \in \mathcal{T}}$ the global forwarding strategy for single-step computations.

To ensure all tasks are fulfilled, every data packet must be eventually forwarded to some computation unit, and every result packet must be delivered to the corresponding destination. 
Specifically, the data flows are either forwarded to nearby nodes or to local computation unit, and the result flows exit the network at the destination.
Therefore, for all $(d,m) \in \mathcal{T}$ and $i\in \mathcal{S}$, it holds that
\begin{equation}
    \sum_{j \in \left\{0\right\} \cup \mathcal{V} } \phi_{ij}^-(d,m) = 1, \quad
    \sum_{j \in \mathcal{V} } \phi_{ij}^+(d,m) = 
    \begin{cases} 
    1, \, \text{if } i \neq d, 
    \\ 0, \, \text{if } i = d.
    \end{cases} 
    \label{FlowConservation}
\end{equation}

\noindent\textbf{Communication cost.} 
We assume the communication cost on link $(i,j)$ is $D_{ij}(F_{ij})$, where $F_{ij}$ is the total flow rate (bit/s) on link $(i,j)$, given by
\begin{equation*}
    \textstyle F_{ij} = \sum\limits_{(d,m) \in S} \left(L_m^-f_{ij}^-(d,m) + L_m^+f_{ij}^+(d,m)\right),
\end{equation*}
and $D_{ij}(\cdot)$ is an increasing, continuously differentiable and convex function. 
 Such convex costs subsume a variety of existing cost functions including commonly adopted linear cost\cite{ioannidis2018jointly}. It also incorporates performance metrics that reflect the network congestion status.
For example, provided that $\mu_{ij}$ is the service rate of an M/M/1 queue \cite{bertsekas2021data} with $F_{ij} < \mu_{ij}$, $D_{ij}(F_{ij}) = {F_{ij}}/\left({\mu_{ij}-F_{ij}}\right)$ gives the average number of packets waiting for or under transmission on link $(i,j)$, and is proportional to the average time for a packet from entering the queue to transmission completion.
One can also approximate the link capacity constraint $F_{ij} \leq C_{ij}$ (e.g., in \cite{liu2019joint}) by a smooth convex function that goes to infinity when approaching the capacity limit $C_{ij}$. 

\vspace{0.5\baselineskip}
\noindent\textbf{Computation cost.}
We define the computation workload at node $i$ as
\vspace{-0.1\baselineskip}
\begin{equation*}
     G_i = \sum_{m =1}^M w_{im} \left(\sum_{d:(d,m)\in\mathcal{T}} g_i(d,m)\right),
\end{equation*}
where $w_{im} > 0$ is the weight for type $m$ at node $i$. 
We assume the computation cost at node $i$ is $C_i(G_i)$, where $C_i(\cdot)$ is an increasing, continuously differentiable and convex function.
For instance, if the computation of type $m$ requires $c_m$ CPU cycles per 
bit of input data.
By setting $w_{im} = c_m$ and $C_i(G_i) = G_i$, computation cost $C_i(G_i)$ measures the total CPU cycles. 
Alternatively, let $v_i^m$ be the computation speed of type $m$ at $i$ and $w_{im} = c_m/ v_i^m$, cost $C_i(G_i)$ measures the CPU runtime.
Function $C_i(G_i)$ can also incorporate computation congestion (e.g., average number of packets waiting for available processor or being served at CPU).
When both $D_{ij}(F_{ij})$ and $C_i(G_i)$ represent queue lengths, by Little's Law, the network aggregated cost is proportional to the expected packet system delay.

Note that for a network with heterogeneous computation resources, our formulation is more flexible than that in \cite{sahni2020multi}\cite{chen2021edgeconomics}, where a task always yields the same computation cost wherever it is performed. Our model captures the fact that in practical edge networks, the workload for a certain task may be very different depending on where it is computed, e.g., some parallelizable tasks are easier at nodes employing GPU acceleration, but slower at others.

\vspace{0.5\baselineskip}
\noindent\textbf{Problem formulation.}  
We aim at minimizing the total cost of links and devices for both communication and computation, 
  \begin{subequations}
\begin{align}
    \min_{\boldsymbol{\phi}} \quad & T(\boldsymbol{\phi}) = \sum_{(i,j) \in \mathcal{E}}D_{ij}(F_{ij}) + \sum_{i \in \mathcal{V}} C_i(G_i) 
    \\ \text{subject to} \quad & 
    \boldsymbol{\phi} 	\in [\boldsymbol{0},\boldsymbol{1}] \text{ and (\ref{FlowConservation}) holds.}  
\end{align}
\label{JointProblem_nodebased}
  \end{subequations}
We do not explicitly impose any link or computation capacity constraints in \eqref{JointProblem_nodebased} since they are already incorporated in the cost functions. 

Note that even for the single-step computations,
problem \eqref{JointProblem_nodebased} is not convex in $\boldsymbol{\phi}$.
Nevertheless, we are able to globally solve \eqref{JointProblem_nodebased} by providing a set of sufficient optimality conditions. 

\subsection{Optimality Conditions for Single-step Computations}
\label{sec:optimality_single}
We first establish a set of KKT necessary conditions for \eqref{JointProblem_nodebased}, and demonstrate by example that such necessary conditions could lead to arbitrarily suboptimal solutions.
Then, we establish a set of sufficient optimality conditions by modifying the KKT condition to overcome its degenerate cases.

\noindent\textbf{Necessary condition.}
We start by giving closed-form derivatives of $T$.
Our analysis follows \cite{gallager1977minimum} and makes non-trivial extensions to data and result flows, as well as in-network computation.
For $(i,j) \in \mathcal{E}$ and $(d,m) \in \mathcal{T}$, the marginal cost for a marginal increase of $\phi_{ij}^-(d,m)$ consists of two components, 1) the marginal communication cost on link $(i,j)$, and 2) the marginal cost of increasing data traffic $t_j^-(d,m)$.
Similarly, the marginal cost of increasing $\phi_{i0}^-(d,m)$ consists of the marginal computation cost at $i$, and the marginal cost of increasing result traffic $t_i^+(d,m)$.
Formally,
\begin{equation}
\begin{aligned}
    &\frac{\partial T}{\partial \phi_{ij}^-(d,m)} =  t_i^-(d,m) \left[L_m^- D_{ij}^\prime(F_{ij}) + \frac{\partial T}{\partial t^-_j(d,m)}\right],
    \\ & \frac{\partial T}{\partial \phi_{i0}^-(d,m)}= t_i^-(d,m) \left[ w_{im} C_i^\prime(G_i) + \frac{\partial T}{\partial t_i^+(d,m)} \right].
\end{aligned}
\label{partial_D_phi-}
\end{equation}
The marginal cost of increasing $\phi_{ij}^+(d,m)$ also consists of two components, i.e., the cost due to increasing flow rate on link $(i,j)$, and the cost due to increasing result traffic $t_j^+(d,m)$ at node $j$. Formally, 
\begin{equation}
     \frac{\partial T}{\partial \phi_{ij}^+(d,m)} = 
     t_i^+(d,m) \left[L_m^+D_{ij}^\prime(F_{ij}) + \frac{\partial T}{\partial t_j^+(d,m)} \right].\label{partial_D_phi+}
\end{equation}
In \eqref{partial_D_phi-}, term ${\partial T}/{\partial t^-_j(d,m)}$ is a weighted sum of marginal costs for out-going links and local computation unit, namely,
\begin{equation}
    \begin{aligned}
         \frac{\partial T}{ \partial t^-_j(d,m)} &= \sum_{j \in \mathcal{N}(i)} \phi_{ij}^-(d,m) \left[ L_m^- D_{ij}^\prime(F_{ij}) + \frac{\partial T}{ \partial t_j^-(d,m)}\right]
    \\ &+ \phi_{i0}^-(d,m) \left[w_{im}C_i^\prime(G_i) + \frac{\partial T}{\partial t_i^+(d,m)}\right].
    \end{aligned}
    \label{partial_D_t-}
\end{equation}
Similarly, the term $\partial T/\partial t_i^+(d,m)$ is given by
\begin{align}
    \frac{\partial T}{\partial t_i^+(d,m)} = \sum_{j \in \mathcal{N}(i)} \phi_{ij}^+(d,m) \left[ L_m^+ D_{ij}^\prime(F_{ij}) +  \frac{\partial T}{\partial t_j^+(d,m)}\right].
    \label{partial_D_t+}
\end{align}
With the absence of computation offloading, to formally derive \eqref{partial_D_phi-} \eqref{partial_D_phi+} and \eqref{partial_D_t-} \eqref{partial_D_t+} in detail, a rigorous proof is presented in Theorem 2 \cite{gallager1977minimum}.
One could calculate $\partial T/\partial t_i^-(d,m)$ and $\partial T/\partial t_i^+(d,m)$ recursively by \eqref{partial_D_t-} and \eqref{partial_D_t+} starting with $i = d$, since the result flow does not introduce any marginal cost at the destination, i.e., $\partial T/\partial t_d^+(d,m) = 0$.

Therefore, for single-step computations, we present a set of KKT necessary conditions for the optimal solution to \eqref{JointProblem_nodebased}, given in Lemma \ref{Lemma_Necessary_single}.
\begin{lem} \label{Lemma_Necessary_single}
Let $\boldsymbol{\phi}$ be the global optimal solution to \eqref{JointProblem_nodebased}, then for all $i \in \mathcal{V}$, $(d,m) \in \mathcal{T}$ and $j\in\left\{0\right\} \cup \mathcal{N}(i)$,
\begin{equation*}
    \frac{\partial T}{ \partial \phi_{ij}^-(d,m)}  
    \begin{cases}
     = \min\limits_{k \in \left\{0\right\} \cup \mathcal{N}(i) } \frac{\partial T}{ \partial \phi_{ik}^-(d,m)} , \quad \text{if } \phi_{ij}^-(d,m) >0,
    \\ \geq \min\limits_{k \in \left\{0\right\} \cup \mathcal{N}(i) } \frac{\partial T}{ \partial \phi_{ik}^-(d,m)} ,  \quad \text{if } \phi_{ij}^-(d,m) =0,
    \end{cases}
\end{equation*}
and for all $j \in \mathcal{N}(i)$,
\begin{equation*}
    \frac{\partial T}{ \partial \phi_{ij}^+(d,m)}  
     \begin{cases}
     = \min\limits_{k \in \mathcal{N}(i) } \frac{\partial T}{ \partial \phi_{ik}^+(d,m)} , \quad \text{if } \phi_{ij}^+(d,m) >0,
    \\ \geq \min\limits_{k \in \mathcal{N}(i) } \frac{\partial T}{ \partial \phi_{ik}^+(d,m)}, \quad \text{if } \phi_{ij}^+(d,m) =0.
    \end{cases}
\end{equation*}
\end{lem}
\begin{proof}
{
Lemma \ref{Lemma_Necessary_single} is a special case of Lemma \ref{Lemma_Necessary_chain} in Section \ref{sec:optimality_chain} for single-step computations. 
}
\end{proof}

\begin{figure}[htbp]
\centerline{\includegraphics[width=0.48\textwidth]{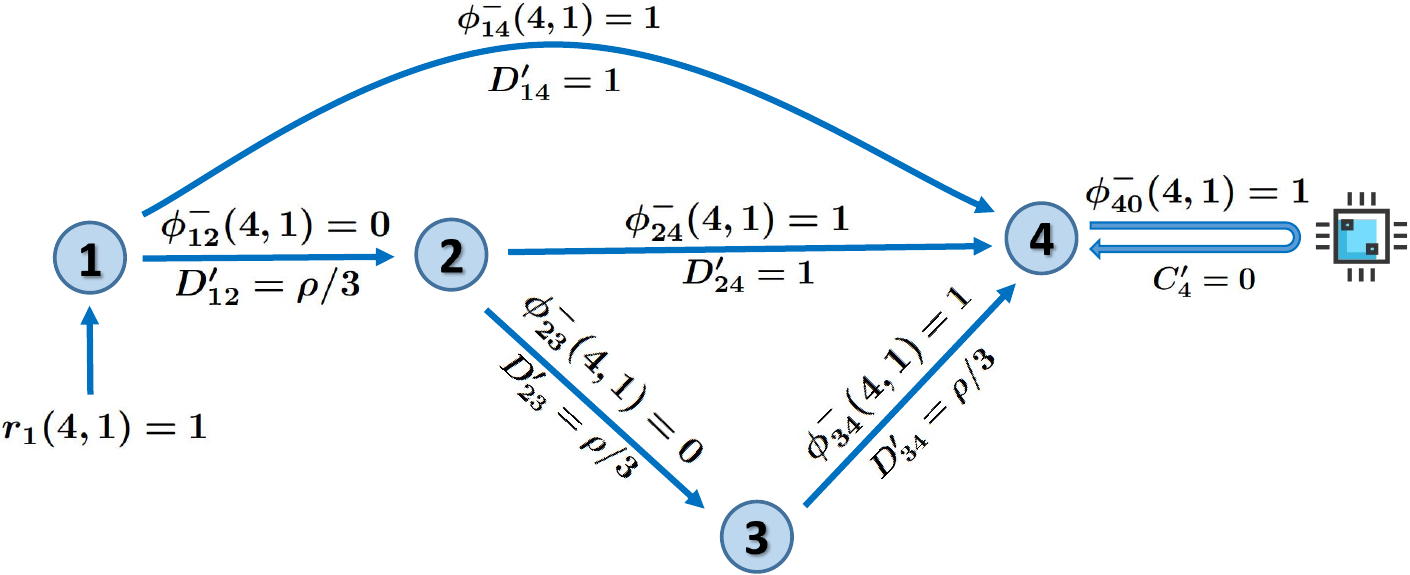}}
\caption{A simple example that KKT conditions in Lemma \ref{Lemma_Necessary_single} leads to arbitrarily suboptimal solutions.}
\label{fig_kkt_suboptimal}
\end{figure}

The KKT conditions given in Lemma \ref{Lemma_Necessary_single} are not sufficient for global optimality \cite{zhang2022optimal,gallager1977minimum}. 
As a matter of fact, forwarding strategies $\boldsymbol{\phi}$ satisfying the KKT conditions may perform arbitrarily worse compared to the global optimal solution.
\begin{prop}
\label{prop_arbitrarily_worse}
   For any $0<\rho<1$, there exists a scenario (i.e., network $\mathcal{G}$, tasks $\mathcal{T}$, cost functions $F_{ij}(\cdot)$, $G_i(\cdot)$, and input rates $\boldsymbol{r}$) such that $\frac{T(\boldsymbol{\phi}^*)}{T(\boldsymbol{\phi})} = \rho$, where $\boldsymbol{\phi}$ is feasible to \eqref{JointProblem_nodebased} and satisfies the condition in Lemma \ref{Lemma_Necessary_single}, and $\boldsymbol{\phi}^*$ is an optimal solution to \eqref{JointProblem_nodebased}.
\end{prop}
\begin{proof}
To see this, we next construct a scenario satisfying that $T(\boldsymbol{\phi}^*)/T(\boldsymbol{\phi}) = \rho$ for an arbitrarily given $ 0<\rho<1$.
Consider the simple example in Fig. \ref{fig_kkt_suboptimal}, where only one task $(d,m)$ exists with $(d,m) = (4,1)$. 
Exogenous input data occurs only at node $1$, and the computation unit is only equipped at node $4$.
For simplicity, we assume all cost functions are linear with their marginals shown in the figure. We focus solely on the communication cost by letting $C^\prime_4 = 0$.
Consider the strategy $\boldsymbol{\phi}$ shown in the figure, where the data flow is routed directly from node $1$ to node $4$, and no traffic goes through node $2$.
It can be easily verified that the condition in Lemma \ref{Lemma_Necessary_single} holds for the given $\boldsymbol{\phi}$, with the total cost $T(\boldsymbol{\phi}) = 1$.
However, consider another forwarding strategy $\boldsymbol{\phi}^*$: the entire data traffic is routed along path $1 \to 2 \to 3 \to 4$, incurring a total cost of $\rho$.
It is evident that $\boldsymbol{\phi}^*$ reaches the global optimum for the given network scenario, implying $T(\boldsymbol{\phi^*})/T(\boldsymbol{\phi}) = \rho$.
The ratio of $T(\boldsymbol{\phi^*})$ and $T(\boldsymbol{\phi})$ can be arbitrarily small as $\rho$ varies, that is, the KKT condition yields arbitrarily suboptimal solutions.
\end{proof}

The underlying intuition is that the KKT condition in Lemma \ref{Lemma_Necessary_single} automatically holds for the $i$ and $(d,m)$ with $t_i^-(d,m) = 0$ or $t_i^+(d,m) = 0$ (i.e., when no data or result traffic of task $(d,m)$ is going through node $i$), regardless of the actual forwarding strategy at node $i$. 

\vspace{0.5\baselineskip}
\noindent\textbf{Sufficient condition.}
We now address the non-sufficiency of the KKT condition.
Inspired by \cite{gallager1977minimum}, we introduce a modification to the KKT condition that leads to a sufficient condition for optimality in \eqref{JointProblem_nodebased}.
Observing that for any $i$ and $(d,m)$, the traffic term $t_i^-(d,m)$ and $t_i^+(d,m)$ repeatedly exists in RHS of \eqref{partial_D_phi-} and \eqref{partial_D_phi+} respectively for all $j \in \{0\} \cup \mathcal{V}$. 
Therefore, following \cite{gallager1977minimum,zhang2022optimal}, we remove the traffic terms in the conditions given by Lemma \ref{Lemma_Necessary_single}, leading to a set of sufficient optimality condition \eqref{Condition_sufficient_single} that globally solves \eqref{JointProblem_nodebased} for single-step computations. 

\begin{theo} \label{Thm_Sufficient_single}
Let $\boldsymbol{\phi}$ be feasible for (\ref{JointProblem_nodebased}).  If for all $i \in \mathcal{V}$, $(d,m) \in \mathcal{T}$ and $j\in\left\{0\right\} \cup \mathcal{N}(i)$, it holds that
\begin{subequations}
\label{Condition_sufficient_single}
\begin{equation}
    \delta_{ij}^-(d,m) \begin{cases}
    = \min\limits_{k \in \left\{0\right\}  \cup \mathcal{N}(i)} \delta_{ik}^-(d,m), \quad \text{if } \phi_{ij}^-(d,m) >0,
    \\ \geq \min\limits_{k \in \left\{0\right\}  \cup \mathcal{N}(i)} \delta_{ik}^-(d,m), \quad \text{if } \phi_{ij}^-(d,m) =0,
    \end{cases}
\end{equation}
and for all $j \in \mathcal{N}(i)$, it holds that
\begin{equation}
     \delta_{ij}^+(d,m) \begin{cases}
    = \min\limits_{k \in \mathcal{N}(i)} \delta_{ik}^+(d,m), \quad \text{if } \phi_{ij}^+(d,m) >0,
    \\ \geq \min\limits_{k \in \mathcal{N}(i)} \delta_{ik}^+(d,m),  \quad \text{if } \phi_{ij}^+(d,m) =0,
    \end{cases}
\end{equation}
\end{subequations}
then $\boldsymbol{\phi}$ is a global optimal solution to \eqref{JointProblem_nodebased},
where 
$\delta_{ij}^-(d,m)$ and $\delta_{ij}^+(d,m)$ are defined as
\begin{equation}
\begin{aligned}
    \delta_{ij}^-(d,m) &= 
    \begin{cases}
    L_m^- D_{ij}^\prime(F_{ij}) + \frac{\partial T}{\partial t_j^-(d,m)}, &\text{if } j \neq 0,
    \\ w_{im} C_i^\prime(G_i) + \frac{\partial T}{\partial t_i^+(d,m)} , & \text{if } j =0,
    \end{cases} 
    \\ \delta_{ij}^+(d,m) &=  L_m^+ D_{ij}^\prime(F_{ij}) +  \frac{\partial T}{\partial t_j^+(d,m)}.
\end{aligned}\label{delta}
\end{equation}
\end{theo}
\begin{proof}
    { 
    Theorem \ref{Thm_Sufficient_single} is a special case of Theorem \ref{Thm_Sufficient_chain} in Section \ref{sec:optimality_chain} for single-step computations. 
    }
\end{proof}

To see the difference between the KKT necessary condition given by Lemma \ref{Lemma_Necessary_single} and the sufficient condition given by Theorem \ref{Thm_Sufficient_single}, consider again the example in Fig. \ref{fig_kkt_suboptimal}.
For any $\boldsymbol{\phi}$ satisfying \eqref{Condition_sufficient_single}, it must hold that $\phi_{12}(4,1) = 1$, $\phi_{23}(4,1) = 1$ and $\phi_{34}(4,1) = 1$, precisely indicating the shortest path $1 \to 2 \to 3 \to 4$ as expected.
Intuitively, condition \eqref{Condition_sufficient_single} suggests that each node handles incremental arrival flow in the way that achieves its minimum marginal cost -- either by forwarding to neighbors, or to its local CPU.

\section{Offloading of Service Chain Applications}
\label{sec:service chain}

In Section \ref{sec:single step}, we formulated the single-step computation offloading problem, and provided a sufficient optimality condition that solves the non-convex optimization problem globally.
In this section, we generalize the formulation and method in Section \ref{sec:single step} to general service chain computation applications.
We first generalize our network model in Section \ref{Section:model_chain}, and present the sufficient optimality condition for forwarding and offloading service chain application in Section \ref{sec:optimality_chain}.
Furthermore, we show in Section \ref{sec:geodesic} the geodesic convexity under additional assumptions, and provide theoretical insights into the sufficiency of our proposed condition.
To distinguish from Section \ref{sec:single step}, our major notations for service chain applications are summarized in Table \ref{table:tab1}.

\subsection{Generalized Formulation for Service Chain Applications}
\label{Section:model_chain}

\noindent\textbf{Applications and tasks.} 
Different from our previous formulation of single-step computation tasks in Section \ref{Section:model_single}, for the general service chain computation offloading, we assume the communication and computation performed in network $\mathcal{G}$ are driven by a finite number of \emph{applications}. 
We denote the set of applications by $\mathcal{A}$. 
Each application consists of a chain of services, i.e., a finite number of \emph{tasks} need to be performed sequentially with a pre-determined order.
Each application has one pre-specified destination, to which the final results of the service chain need to be delivered. 
For application $a \in \mathcal{A}$, we denote by $\mathcal{T}_a$ the (ordered) set of tasks affiliated to $a$, by $d_a \in \mathcal{V}$ the destination node for final results, and by task $t_{a,k} \in \mathcal{T}_a$ the $k$-th task of this service chain. 
We also assume  $\mathcal{T}_a \bigcap \mathcal{T}_b = \emptyset$ for any $a,b \in \mathcal{A}, a \neq b$, and denote by $\mathcal{T} = \bigcup_{a \in \mathcal{A}}\mathcal{T}_a$ the set of all tasks. 
Note that for model simplicity, we do not use the concept of \emph{computation type} for the tasks of service chain applications, as we inherently assume every task is of a different type.

\vspace{0.5\baselineskip}
\noindent\textbf{Flow conservation.}
Similar to Section \ref{Section:model_single}, we adopt a flow model to estimate the long-term averaged network behavior. 
Flows are injected into the network at data sources, and exit the network at application destinations after being processed for the corresponding chain of tasks.

The data input rate for application $a$ at node $i$ is $r_i(a)$ (packet/sec), where each data packet has size $L_{(a,0)}$ (bit).
We define $\boldsymbol{r} = [r_i(a)]_{i \in \mathcal{V}, a \in \mathcal{A}}$ as the global input vector.
For application $a$, the data flows are injected into the network from nodes $i$ with $r_i(a) >0$, forwarded in a hop-by-hop manner to nodes that decide to perform the first task of $a$. After the first task, data flows are converted into flows of intermediate results, and are further forwarded for the next task in the service chain. Eventually, final results are delivered to the destination $d_a$.
We assume the flows of application $a \in \mathcal{A}$ are categorized into $|\mathcal{T}_a|+1$ \emph{stages}, where stage $(a,k)$ with $k = 0, 1, \cdots, |\mathcal{T}_a|$ represents the (intermediate) results that have finished the $k$-th computation task of $a$. Particularly, we say the data flows are of stage $(a,0)$, and the final results are of stage $(a,|\mathcal{T}_a|)$. 
We assume packets of stage $(a,k)$ are of size $L_{(a,k)}$ (bit), and denote the set of all stages by $\mathcal{S} = \left\{(a,k)\big|a \in \mathcal{A}, k = 0,1\cdots,\left|\mathcal{T}_a\right|\right\}$.

Generalizing Section \ref{Section:model_single}, we let $t_i(a,k)$ denote the \emph{traffic} (packet/s) of stage $(a,k) \in \mathcal{S}$ at node $i$. 
Specifically, $t_i(a,k)$ includes both the packet rate of stage $(a,k)$ that is generated at $i$ (this can be the input data rate $r_i(a)$ if $k = 0$, or the newly generated intermediate results at $i$'s computation unit if $k \neq 0$), and the packet rate that is forwarded from other nodes to $i$.
We let $\phi_{ij}(a,k) \in [0,1]$ be the \emph{forwarding} variable for $i,j \in \mathcal{V}$ and $(a,k) \in \mathcal{S}$, i.e., of the traffic $t_i(a,k)$ that arrive at $i$, node $i$ forwards a fraction of $\phi_{ij}(a,k)$ to node $j$.
Moreover, if $k \neq |\mathcal{T}_a|$, node $i$ forwards a fraction $\phi_{i0}(a,k)$ of $t_{i}(a,k)$ to its local CPU to perform computation of the $k+1$-th task.

We assume that for every data or intermediate result packet forwarded to the CPU for computation, one next-stage packet is generated accordingly.
Therefore, for $k = 0$, 
\begin{equation*}
    t_i(a,0) = \sum\nolimits_{j \in \mathcal{V}} t_{j}(a,0) \phi_{ji}(a,0) + r_i(a),
\end{equation*}
 and for $k \neq 0$,
\begin{equation*}
    t_i(a,k) = \sum\nolimits_{j \in \mathcal{V}} t_{j}(a,k) \phi_{ji}(a,k) + t_{i}(a,k-1) \phi_{i0}(a,k-1).
\end{equation*}

\begin{table}[t]
\footnotesize
\begin{tabular}{l | l }
\hline
$\mathcal{G} = (\mathcal{V},\mathcal{E})$ & Network graph $\mathcal{G}$, set of nodes $\mathcal{V}$ and links $\mathcal{E}$\\
$\mathcal{A}$ & Set of computation applications
\\
$\mathcal{T}_a, d_a$ & Service chain tasks and destination of application $a \in \mathcal{A}$ \\
$\mathcal{S}$ & Set of all stages $(a,k)$ where $a \in \mathcal{A}$, $k = 0,\cdots,|\mathcal{T}_a|$\\
$L_{(a,k)}$ & Packet size of stage $(a,k) \in \mathcal{S}$\\
$r_i(a)$ & Exogenous input data rate for application $a$ at node $i$ \\
$t_{i}(a,k)$ & Traffic of stage $(a,k)$ at node $i$ \\
$\phi_{ij}(a,k)$ & Fraction of $t_{i}(a,k)$ forwarded to node $j$ (if $j \neq 0$) \\
$\phi_{i0}(a,k)$ & Fraction of $t_{i}(a,k)$ assigned to CPU at $i$ \\
$f_{ij}(a,k)$ & Rate (packet/sec) of stage $(a,k)$ on link $(i,j)$\\
$g_i(a,k)$ & Rate (packet/sec) of stage $(a,k)$ assigned to CPU at $i$\\
$D_{ij}(F_{ij})$ & Transmission cost (e.g. queueing delay) on link $(i,j)$\\
$C_i(G_i)$ & Computation cost (e.g. CPU runtime) at node $i$ \\
\hline
\end{tabular}
\caption{Major notations for service chain applications}
\label{table:tab1}
\end{table}

Let $\boldsymbol{\phi} = [ \phi_{ij}(a,k)]_{(a,k)\in \mathcal{S}, i \in \mathcal{V}, j \in \{0\} \cup \mathcal{V}}$ be the \emph{global forwarding strategy} for service chain applications. The flow conservation \eqref{FlowConservation} is replaced with the following constraint,
\begin{align}
    \sum_{j \in \{0\} \cup \mathcal{V}} \phi_{ij}(a,k) = \begin{cases}
        0, &\quad \text{if } k = \left|\mathcal{T}_a\right| \text{ and } i = d_a,
        \\ 1, &\quad \text{otherwise}.
    \end{cases}\label{FlowConservation_phi}
\end{align}
Conservation \eqref{FlowConservation_phi} guarantees that all service chain applications are fully accomplished in the network, i.e., every data packet that enters the network is eventually converted to a result packet and exits the network at the corresponding destination.

\begin{figure}[t!]
\centerline{
\includegraphics[width=0.48\textwidth]{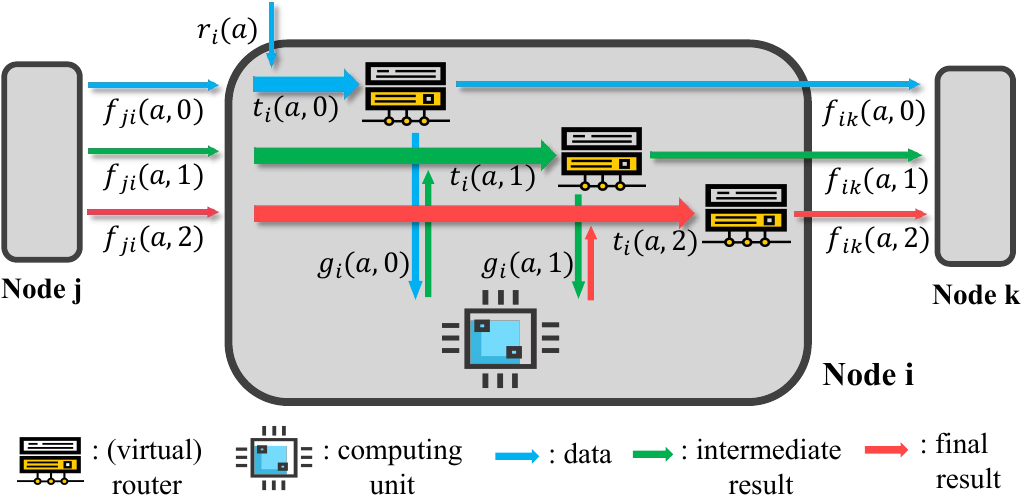}
}
\caption{Illustration of flows for nodes $j \to i \to k$ for a service chain computation application} 
\label{fig_node_behavior}
\end{figure}

Let $f_{ij}(a,k)$ be the rate (packet/sec) for stage $(a,k)$ transmitted on link $(i,j)$, and let $g_i(a,k)$ be the rate (packet/sec) for stage $(a,k)$ forwarded to $i$'s CPU for computation. Then,
\begin{equation*}
    f_{ij}(a,k) = t_{i}(a,k) \phi_{ij}(a,k), \quad g_{i}(a,k) = t_{i}(a,k) \phi_{i0}(a,k).
\end{equation*}
The {total flow rate} (bit/sec) on link $(i,j)\in\mathcal{E}$ is given by
\begin{equation*}
    F_{ij} = \sum\nolimits_{(a,k) \in \mathcal{S}} L_{(a,k)} f_{ij}(a,k),
\end{equation*}
and the total computation workload at node $i \in \mathcal{V}$ is given by
\begin{equation*}
    G_i = \sum\nolimits_{ (a,k) \in \mathcal{S}} w_i(a,k) g_i(a,k),
\end{equation*}
where $w_i(a,k)$ is the computation workload for node $i$ to perform the $(k+1)$-th task of application $a$ on a single input packet. 
Fig. \ref{fig_node_behavior} illustrates the service chain flow model.
Similar to Section \ref{Section:model_single}, we denote by $D_{ij}(F_{ij})$ the communication cost on link $(i,j)$, and $C_i(G_i)$ the computation cost at node $i$, and assume they are increasing convex. 

\vspace{0.5\baselineskip}
\noindent\textbf{Problem Formulation.}
For analytical simplicity, we assume the network and application pattern are quasi-static. 
For given network topology and cost functions, we denote the region of feasible link flows and workloads by
\begin{align*}
    \mathcal{D}_{F} = \bigg\{ F_{ij}, G_i \Big| &D_{ij}(F_{ij}) < \infty, \, \forall  (i,j) \in \mathcal{E}  \\&\text{ and } C_i(G_i) < \infty,\, \forall i \in \mathcal{V}\bigg\}.
\end{align*}

For a fixed network and application pattern, let $\boldsymbol{F}(\boldsymbol{r},\boldsymbol{\phi}) = ([F_{ij}]_{(i,j)\in\mathcal{E}},[G_i]_{i \in \mathcal{V}})$ be the link flows and computation workloads given input $\boldsymbol{r}$ and forwarding strategy $\boldsymbol{\phi}$.
We denote the feasible region of the forwarding strategy by
\begin{align*}
    \mathcal{D}_{\boldsymbol{\phi}}(\boldsymbol{r}) = \left\{\boldsymbol{\phi}\Big| \text{ \eqref{FlowConservation_phi} holds, and } \boldsymbol{F}(\boldsymbol{r},\boldsymbol{\phi}) \in \mathcal{D}_{F}\right\}.
\end{align*}
The stable input rate region is given by
\begin{align*}
    \mathcal{D}_{\boldsymbol{r}} = \left\{ \boldsymbol{r} \geq \boldsymbol{0} \Big| \mathcal{D}_{\boldsymbol{\phi}}(\boldsymbol{r}) \neq \emptyset \right\}.
\end{align*}
We assume $\mathcal{D}_{\boldsymbol{r}}$ has non-empty interior, and $\boldsymbol{r} \in \mathcal{D}_{\boldsymbol{r}}$.
Without ambiguity, we abbreviate $\mathcal{D}_{\boldsymbol{\phi}}(\boldsymbol{r})$ by $\mathcal{D}_{\boldsymbol{\phi}}$.

To optimally embed service chain applications into $\mathcal{G}$, we minimize the aggregated transmission and computation cost in the network for all data and (intermediate) results,
\begin{subequations}\label{ServiceChainObj}
\begin{align}
&\min_{\boldsymbol{\phi}} \quad T(\boldsymbol{\phi}) = \sum\nolimits_{(i,j)\in \mathcal{E}} D_{ij}(F_{ij}) + \sum\nolimits_{i \in \mathcal{V}} C_{i}(G_i), 
    \\ &\text{subject to} \quad \boldsymbol{\phi} \in \mathcal{D}_{\boldsymbol{\phi}}.
\end{align}
\end{subequations}

\subsection{Optimality Conditions for Service Chain Applications}
\label{sec:optimality_chain}

To tackle \eqref{ServiceChainObj}, we generalize our method in Section \ref{sec:optimality_single} to service chain applications.
Specifically, we first present a set of KKT necessary conditions for \eqref{ServiceChainObj}, and provide a set of sufficient optimality conditions that solves \eqref{ServiceChainObj} globally.

\vspace{0.5\baselineskip}
\noindent\textbf{KKT necessary condition.}
To generalize the KKT condition given by Lemma \ref{Lemma_Necessary_single} to service chain applications, we rewrite the closed-form derivatives of the aggregated transmission and computation cost $T$ in Section~\ref{sec:optimality_single}. 
The detailed explanation is almost a replica of that for \eqref{partial_D_phi-} \eqref{partial_D_phi+}, and is omitted here.
Formally, for $(i,j) \in \mathcal{E}$,
\begin{subequations}\label{partial_D_phi}
\begin{equation}
    \frac{\partial T}{\partial \phi_{ij}(a,k)} =  t_i(a,k) \left(L_{(a,k)} D^\prime_{ij}(F_{ij}) + \frac{\partial T}{\partial t_j(a,k)}\right),
\label{partial_D_phi_j}
\end{equation}
for $j = 0$ and $k\neq |\mathcal{T}_a|$,
\begin{equation}
    \frac{\partial T}{\partial \phi_{i0}(a,k)} =  t_i(a,k)\left(w_i(a,k) C^\prime_{i}(G_{i}) + \frac{\partial T}{\partial t_i(a,k+1)}\right). 
\label{partial_D_phi_0}
\end{equation}
\end{subequations}

In \eqref{partial_D_phi}, if $k \neq |\mathcal{T}_a|$,
\begin{subequations}
\label{partial_D_r}
\begin{align}
    &\frac{\partial T}{\partial t_i(a,k)} = \phi_{i0}(a,k)\left(w_i(a,k) C^\prime_{i}(G_{i}) + \frac{\partial T}{\partial t_i(a,k+1)}\right)\nonumber
   \\ &+ \sum_{j \in \mathcal{V} } \phi_{ij}(a,k)\left(L_{(a,k)} D^\prime_{ij}(F_{ij}) + \frac{\partial T}{\partial t_j(a,k)}\right),\label{partial_D_r_1}
\end{align}
and for $k = |\mathcal{T}_a|$, recall $\phi_{i0}(a,|\mathcal{T}_a|) \equiv 0$, it holds that
\begin{equation}
\begin{aligned}
\label{partial_D_r_2}
    \frac{\partial T}{\partial t_i(a,|\mathcal{T}_a|)} & = 
    \sum_{j \in \mathcal{V} } \phi_{ij}(a,|\mathcal{T}_a|) \times
    \\& \left(L_{(a,|\mathcal{T}_a|)} D^\prime_{ij}(F_{ij}) + \frac{\partial T}{\partial t_j(a,|\mathcal{T}_a|)}\right),
\end{aligned}
\end{equation}
\end{subequations}
where $\partial T/\partial t_i(a,k)$ can be recursively computed by \eqref{partial_D_r}.\footnote{For coherence, we let $\partial T/\partial \phi_{ij}(a,k) \equiv \infty$ for $(i,j) \not \in \mathcal{E}$, and $\partial T/\partial \phi_{i0}(a,|\mathcal{T}_a|) \equiv \infty$ as the final results will not be further computed.}
Lemma \ref{Lemma_Necessary_chain} gives the KKT necessary condition of \eqref{ServiceChainObj} for service chain applications.
\begin{lem} 
\label{Lemma_Necessary_chain}
Let $\boldsymbol{\phi}$ be an optimal solution to \eqref{ServiceChainObj}, then for all $i \in \mathcal{V}$, $j\in\left\{0\right\} \cup \mathcal{V}$, and $(a,k)\in \mathcal{S}$,
\begin{equation}
    \frac{\partial T}{ \partial \phi_{ij}(a,k)}  
    \begin{cases}
     = \min\limits_{j^\prime \in \left\{0\right\} \cup \mathcal{V} } \frac{\partial T}{ \partial \phi_{ij^\prime}(a,k)} , \, \text{if } \phi_{ij}(a,k) >0,
    \\ \geq \min\limits_{j^\prime \in \left\{0\right\} \cup \mathcal{V} } \frac{\partial T}{ \partial \phi_{ij^\prime}(a,k)} ,  \, \text{if } \phi_{ij}(a,k) =0.
    \end{cases}
\label{Condition_KKT_chain}
\end{equation}
\end{lem}
\begin{proof}
   {Please see Appendix \ref{Proof:Lem_Necessary_chain}}.
\end{proof}

Extending Proposition \ref{prop_arbitrarily_worse}, the KKT condition provided in Lemma \ref{Lemma_Necessary_chain} can lead to arbitrarily worse performance compared to the global optimal solution of \eqref{ServiceChainObj}.
By adopting the modification in Section \ref{sec:optimality_single}, i.e., removing traffic term $t_i(a,k)$ on both sides of \eqref{Condition_KKT_chain}, a set of sufficient optimality conditions for service chain applications are provided in Theorem \ref{Thm_Sufficient_chain}.

\begin{theo}
\label{Thm_Sufficient_chain}
Let $\boldsymbol{\phi}$ be feasible to \eqref{ServiceChainObj}. Then $\boldsymbol{\phi}$ is a global optimal solution to \eqref{ServiceChainObj} if the following holds for all $i \in \mathcal{V}$, $j\in\left\{0\right\} \cup \mathcal{V}$ and $(a,k) \in \mathcal{S}$,
\begin{equation}
    \delta_{ij}(a,k) \begin{cases}
    = \min\limits_{j^\prime \in \left\{0\right\}  \cup \mathcal{V}} \delta_{ij^\prime}(a,k), \, \text{if } \phi_{ij}(a,k) >0,
    \\ \geq \min\limits_{j^\prime \in \left\{0\right\}  \cup \mathcal{V}} \delta_{ij^\prime}(a,k), \, \text{if } \phi_{ij}(a,k) =0,
    \end{cases}
\label{Condition_sufficient_chain}
\end{equation}

where $\delta_{ij}(a,k)$ is the ``modified marginal", given by
\begin{equation}
    \delta_{ij}(a,k) = \begin{cases}
        L_{(a,k)} D^\prime_{ij}(F_{ij}) + \frac{\partial T}{\partial t_j(a,k)}, \quad &\text{if } j \neq 0,
        \\ w_i(a,k) C^\prime_{i}(G_{i}) + \frac{\partial T}{\partial t_i(a,k+1)}, \quad &\text{if } j = 0.
    \end{cases}
\label{delta}
\end{equation}
\end{theo}
\begin{proof}
   Please see Appendix \ref{Proof:Thm_sufficient_chain}.
\end{proof}

\subsection{Perspective From Geodesic}
\label{sec:geodesic}
We remark that although problem \eqref{ServiceChainObj} is non-convex in $\boldsymbol{\phi}$, we manage to globally solve it by rigorously proving the sufficiency of condition \eqref{Condition_sufficient_chain}.
To provide the underlying theoretical insight of such sufficiency, we next show that the objective function $T(\boldsymbol{\phi})$ is geodesically convex in $\boldsymbol{\phi}$ under additional assumptions.
{

Note that condition \eqref{Condition_KKT_chain} and \eqref{Condition_sufficient_chain} coincide at node $i$ and stage $(a,k)$ with strictly positive $t_i(a,k)$.
Therefore, if $\boldsymbol{t} > \boldsymbol{0}$, i.e., $t_i(a,k) > 0$ for all $i$ and $(a,k)$, condition \eqref{Condition_KKT_chain} and \eqref{Condition_sufficient_chain} are essentially equivalent, granting \eqref{Condition_KKT_chain} the global optimality.

\begin{prop}
\label{prop:positive_input}
    If $\boldsymbol{t} > \boldsymbol{0}$ and $\boldsymbol{\phi}$ satisifies \eqref{Condition_KKT_chain}, then $\boldsymbol{\phi}$ optimally solves \eqref{ServiceChainObj}.
\end{prop}

Proposition \ref{prop:positive_input} implies that if $\boldsymbol{t} > \boldsymbol{0}$, the KKT condition \eqref{Condition_KKT_chain} is necessary and sufficient for global optimality of a non-convex problem.
Therefore, we expect a stronger mathematical structure of problem \eqref{ServiceChainObj} beyond the general non-convexity, which has not been discussed by previous works adopting similar approaches \cite{gallager1977minimum,xi2008node,zhang2022optimal}.

The concept of \emph{geodesic convex} is a natural generalization of convexity for sets and functions to Riemannian manifolds~\cite{boumal2023introduction,zhang2016first}.
In this paper, we focus solely on the case when the Riemannian manifold is a Euclidean space.\footnote{Please see Riemannian optimization textbooks, e.g., \cite{boumal2023introduction,zhang2016first}, for detailed definitions and optimization techniques of general geodesic convex functions.}
\begin{defi}[Geodesic convexity on Euclidean space]

Let $C \subset \mathbbm{R}^n$ be a compact convex set.
Function $f: C \to \mathbbm{R}$ is geodesically convex if for any $x_1, x_2 \in C$, there exists a geodesic $\gamma_{x_1x_2}(t)$ joining $x_1$ and $x_2$ with $\gamma_{x_1x_2}(t) \in C$ for $t \in [0,1]$, and $f(\gamma_{x_1x_2}(t))$ is convex with respect to $t$.
\end{defi}

To see the geodesic convexity of $T$ when $\boldsymbol{t} > 0$, for all $(a,k) \in \mathcal{S}$, consider the feasible set $\mathcal{D}_{\boldsymbol{f}}$ of $f_{ij}(a,k)$ given by
\begin{equation*}
\begin{aligned}
    \mathcal{D}_{\boldsymbol{f}} = \left\{[f_{ij}(a,k)]_{(i,j)\in\mathcal{E}, (a,k)\in\mathcal{S}}\bigg| \text{\eqref{FlowConservation_phi} holds } \right\}.
\end{aligned}
\end{equation*}

Suppose $\boldsymbol{t} > \boldsymbol{0}$, then there exists a one-to-one mapping between set $\mathcal{D}_{\boldsymbol{\phi}}$ and $\mathcal{D}_{\boldsymbol{f}}$, where we denote the mapping $\boldsymbol{\phi} \to \boldsymbol{f}$ as $\boldsymbol{f}(\boldsymbol{\phi})$, and the mapping $\boldsymbol{f} \to \boldsymbol{\phi}$ as $\boldsymbol{\phi}(\boldsymbol{f})$.
Specifically, $\phi_{ij}(a,k) = f_{ij}(a,k)/t_i(a,k)$, and
\begin{equation*}
    f_{ij}(a,k) = \sum_{v \in \mathcal{V}} r_v(a)\sum_{p \in \mathcal{P}_{vi}}\prod_{l = 1}^{|p|-1}\phi_{p_l p_{l+1}}(a,k_p(l))
\end{equation*}
where $\mathcal{P}_{vi}$ is the set of \emph{extended paths}\footnote{A \emph{path} $p$ of stage $(a,k)$ refers to a sequence of nodes $(p_1, p_2, \cdots, p_{|p|})$ with $(p_l ,p_{l+1})\in\mathcal{E}$ and $\phi_{p_l p_{l+1}}(a,k)>0$. \emph{Extended path} generalizes the concept of path where $p_{l+1} = p_l$ is allowed if $\phi_{p_l 0} >0$ to represent computation steps. When $p_{l+1} = p_l$, we use $\phi_{p_{l}p_{l+1}}$ to represent $\phi_{p_l 0}$. 
} from node $v$ of stage $(a,0)$ to node $i$ of stage $(a,k)$, and $k_p(l)$ is the number of computation steps on the extended path $p$ before position $l$.

Due to the convexity of $D_{ij}(\cdot)$ and $C_i(\cdot)$, we know $T$ is convex in $\boldsymbol{f}$.
Therefore, the total cost $T$ is a geodesic convex function of $\boldsymbol{\phi}$.

\begin{prop}
    Suppose $\boldsymbol{t} > \boldsymbol{0}$, Then $T(\boldsymbol{\phi})$ is geodesically convex in $\boldsymbol{\phi}$ with the geodesic function 
    \begin{equation*}
        \gamma_{\phi_1\phi_2}(t) = \boldsymbol{\phi}\left((1-t)\boldsymbol{f}(\boldsymbol{\phi}_1) + t \boldsymbol{f}(\boldsymbol{\phi}_2)\right).
    \end{equation*}
\end{prop}

When minimizing a convex function subject to linear inequality constraints, KKT conditions are necessary and sufficient for global optimality (see Proposition 4.4.1 in \cite{bertsekas1997nonlinear}).
When extended to general convex constraints, Slater's Constraint Qualification is required, i.e., there must exist a feasible point that satisfies all inequality constraints strictly (see Proposition 4.3.9 in \cite{bertsekas1997nonlinear}).
By substituting $\phi_{i0}(a,k)$ with $1 - \sum_{j \in \mathcal{V}}\phi_{ij}(a,k)$, problem \eqref{ServiceChainObj} satisfies the Slater's Qualification.
Recently, the sufficiency of KKT condition given Slater's Qualification is extended to Remainnian optimization \cite{jana2020convex}.
Therefore, when $\boldsymbol{t} > \boldsymbol{0}$, condition \eqref{Condition_KKT_chain} itself is sufficient for optimality of \eqref{ServiceChainObj} without requiring condition \eqref{Condition_sufficient_chain} and Theorem \ref{Thm_Sufficient_chain}.
On the other hand, when $\boldsymbol{t} > \boldsymbol{0}$ does not hold, the geodesic convexity of \eqref{ServiceChainObj} no longer holds due to the existence of reflection points when $t_i(a,k) = 0$, and the previous modification technique in Theorem \ref{Thm_Sufficient_chain} can be adopted to eliminate the degenerate cases at these reflection points \cite{gallager1977minimum}.


\subsection{Stability}
By modifying the KKT necessary condition \eqref{Condition_KKT_chain} of problem \eqref{ServiceChainObj}, the sufficient optimality condition \eqref{Condition_sufficient_chain} is obtained.
However, we remark that \eqref{Condition_sufficient_chain} is not a necessary and sufficient condition for optimality.
In fact, the set of $\boldsymbol{\phi}$ satisfying condition \eqref{Condition_sufficient_chain} is a strict subset to that of optimal solutions to \eqref{ServiceChainObj}.
To see this, consider the case when $\boldsymbol{\phi}$ satisfies \eqref{Condition_sufficient_chain} with $t_i(a,k) = 0$ for some $i$ and $(a,k)$.
In such scenario, forwarding variables $[\phi_{ij}(a,k)]_{j \in \{0\}\cup\mathcal{V}}$ can be arbitrarily adjusted without affecting the optimality.


Nevertheless, compared to operating at an arbitrary optimal solution of \eqref{ServiceChainObj}, the network operator may still wish to operate at $\boldsymbol{\phi}$ satisfying condition \eqref{Condition_sufficient_chain}.
Recall that in practical implementation, input rate $\boldsymbol{r}$ is estimated via measuring network status, and can be (slowly) time-varying.
Therefore, the network operator should select a strategy (or set of strategies) that is continuously adjustable according to the time-varying input rate $\boldsymbol{r}$.
The set of $\boldsymbol{\phi}$ characterized by condition \eqref{Condition_sufficient_chain} satisfies the above requirement, formally stated in Theorem \ref{thm:stability}.

\begin{theo}
\label{thm:stability}
Let $\boldsymbol{r} \in \mathcal{D}_{\boldsymbol{r}}$ and $\boldsymbol{\phi} \in \mathcal{D}_{\boldsymbol{\phi}}(\boldsymbol{r})$ satisfies condition \eqref{Condition_sufficient_chain} given input $\boldsymbol{r}$, then there exist a function $\epsilon(\delta)$ for $\delta > 0$, 
such that $\lim_{\delta \to 0}\epsilon(\delta) = 0$, and the following holds:

For all $\Delta \boldsymbol{r}$ that $\left|\Delta \boldsymbol{r}\right| < \delta$ and $\left(\boldsymbol{r} + \Delta \boldsymbol{r}\right) \in \mathcal{D}_{\boldsymbol{r}}$, there exists a corresponding $\Delta\boldsymbol{\phi}$ that $\left|\Delta\boldsymbol{\phi}\right| < \epsilon(\delta)$, $\left(\boldsymbol{\phi} + \Delta\boldsymbol{\phi}\right) \in \mathcal{D}_{\boldsymbol{\phi}}\left(\boldsymbol{r} + \Delta \boldsymbol{r}\right)$, and $\left(\boldsymbol{\phi} + \Delta\boldsymbol{\phi}\right)$ satisfies condition \eqref{Condition_sufficient_chain} given input $\left(\boldsymbol{r} + \Delta \boldsymbol{r}\right)$.
\end{theo}

We defer the proof of Theorem \ref{thm:stability} to Section \ref{sec:convergence}. 
We remark that the requirement in Theorem \ref{thm:stability} need not hold for an arbitrary optimal solution to \eqref{ServiceChainObj}.
To see this, consider the case where the optimal solution yields $t_i(a,0) = 0$ for some $i$ and $(a,k)$.
In such case, variables $\phi_{ij}(a,0)$, $j \in \mathcal{V}$ essentially do not affect the total cost $T$ at all, and thus can be arbitrarily set without violating the optimality.
However, when $r_i(a)$ is increased by $\Delta r > 0$, variables $\phi_{ij}(a,0)$ will need to be completely retuned to accommodate the arrival traffic.
In contrast, condition \eqref{Condition_sufficient_chain} restricts the forwarding variables for $i$ and $(a,k)$ even without the presence of actual traffic $t_i(a,k)$.
}

\section{Distributed Algorithm}
\label{Section:algorithm}
In this section, we propose a distributed algorithm that converges to the global optimal solution of \eqref{ServiceChainObj} specified by condition \eqref{Condition_sufficient_chain}. 
The proposed algorithm is a variant of gradient projection, and is adaptive to moderate changes in exogenous input rates and network topology.
Our method is based on \cite{zhang2022optimal} and generalizes to service chain computation applications.

\subsection{Algorithm overview}
{The existence of routing loops generates redundant flow circulation, wastes network resources, and causes potential instability. 
Therefore, we consider strategy $\boldsymbol{\phi}$ with \emph{loop-free} property.
Specifically, we say there is a \emph{path} of stage $(a,k)$ from node $i$ to node $j$ if there is a sequence of nodes $n_1, \cdots ,n_L$ for which $(n_l,n_{l+1}) \in \mathcal{E}$ and $\phi_{n_l n_{l+1}}(a,k) > 0$ for $l = 1,\cdots,L-1$, with $n_1 = i$ and $n_L = j$.
We say $\boldsymbol{\phi}$ has a \emph{loop} of stage $(a,k)$ if there exists $i$, $j \in \mathcal{V}$, such that $i$ has a path of stage $(a,k)$ to $j$, and vice versa.
We say strategy $\boldsymbol{\phi}$ is \emph{loop-free} if no loops are formed for any stage $(a,k)$ with $a \in \mathcal{A}$, $k = 1,\cdots,|\mathcal{T}_a|$.\footnote{We allow loops concatenated by paths of different stages for the same application, e.g., scenarios where the result destination is the data source.}  
}

We partition time into slots of duration $T_{\text{slot}}$, and let the forwarding strategy of node $i$ during $t$-th slot be $\boldsymbol{\phi}_i^t$.
We assume the network starts with a feasible and loop-free strategy ${\boldsymbol{\phi}}^0$, where the initial total cost $T^0$ is finite.
Let
\begin{equation}
     \boldsymbol{\phi}_i^{t+1} = \boldsymbol{\phi}_i^{t} + \Delta\boldsymbol{\phi}_i^{t},
\end{equation}
The update vector $\Delta\boldsymbol{\phi}_i^{t}$ is calculated by
\begin{equation}
\begin{aligned}
    &\Delta\phi_{ij}^{t}(a,k) =
\\    &\begin{cases}
    - \phi_{ij}^{t}(a,k), \quad \text{ if } j \in \mathcal{B}_i^t(a,k)
    \\ -\min\left\{ \phi_{ij}^t(a,k), \alpha e_{ij}^t(a,k) \right\}, \text{ if } j \not\in \mathcal{B}_i^t(a,k) \text{, } e_{ij}^t(a,k) > 0
    \\ S_i^t(a,k) / N_i^t(a,k),
    \quad \text{ if } j \not\in \mathcal{B}_i^t(a,k) \text{, } e_{ij}^t(a,k) = 0
    \end{cases}
\end{aligned}
\label{dphi_and_dy}
\end{equation}
where $\mathcal{B}_i^t(a,k)$ is the set of \emph{blocked nodes} to suppress routing loops\footnote{Note that forwarding to CPU will never incur loops, i.e., $0 \not \in \mathcal{B}_i^t(a,k)$.}, $\alpha$ is the stepsize, and
\begin{equation}
\begin{aligned}
     & e_{ij}^t(a,k) = \delta_{ij}^t(a,k) - \delta_i^t(a,k), \quad \forall j \not\in \mathcal{B}_i^t(a,k),
     \\ & N_i^t(a,k) = \bigg|\left\{j \not\in \mathcal{B}_i^t(a, k) \big| e_{ij}^t(a,k) = 0 \right\}\bigg|,
     \\ & S_i^t(a, k)= \sum\nolimits_{j \in \mathcal{N}(i) \backslash \mathcal{B}_i^t(a, k) \,:\, e_{ij}^t(a,k) > 0}\Delta\phi_{ij}^{t}(a, k). 
\end{aligned}    
\label{algorithm_detail_eNS}
\end{equation}

The underlying idea of \eqref{dphi_and_dy}\eqref{algorithm_detail_eNS} is to transfer forwarding fractions from non-minimum-marginal directions to the minimum-marginal ones.
$\delta_{ij}^t(a,k)$ and $\delta_{i0}^t(a,k)$ are calculated as in \eqref{delta}, and $\delta_i^t(a,k)$ is given by
\begin{equation}
    \delta_{i}^t(a,k) = \min\nolimits_{j \not\in \mathcal{B}_i^t(a, k)} \delta_{ij}^t(a,k).
\label{delta_ik_t}
\end{equation}

In each update slot, to calculate the modified marginal $\delta_{ij}^t(a,k)$ given by \eqref{delta}, the value $\partial T/\partial t_i(a,k)$ is updated throughout the network with a marginal cost broadcasting mechanism. 
The proposed algorithm is summarized in Algorithm \ref{alg_GP}.
Next, we discuss the details of marginal cost broadcasting and the blocked node set $\mathcal{B}_i^t(a,k)$.

\begin{algorithm}[t]
\SetKwRepeat{DoFor}{do}{for}
\SetKwRepeat{DoDuring}{do}{during}
\SetKwRepeat{DoAt}{do}{at}
\SetKwRepeat{DoWhen}{do}{when}
\SetKwInput{KwInput}{Input}
\KwInput{Feasible and loop-free $\boldsymbol{\phi}^0$, stepsize $\alpha$}
Start with $t=0$.\\
\DoAt{ the end of $t$-th slot}
{
Each node updates $\partial T/\partial t_i(a,k)$ for all $(a, k)$ via the marginal cost broadcasting mechanism in Section \ref{sec:broadcast}.\\
Each node calculates \eqref{algorithm_detail_eNS}.\\
Each node updates strategies $\boldsymbol{\phi}_i^t$ by \eqref{dphi_and_dy}.\\
}
\caption{Gradient Projection (GP)}
\label{alg_GP}
\end{algorithm}

\subsection{Marginal Cost Broadcasting}
\label{sec:broadcast}
We generalize the broadcasting mechanism in \cite{zhang2022optimal} to service chain applications.
Recall from \eqref{delta} that to calculate ${\delta}_{ij}(a,k)$, node $i$ needs $D^\prime_{ij}(F_{ij})$, $C^\prime_i(G_i)$, and the marginal costs due to traffic term, i.e., $\partial T/ \partial t_j(a,k)$ and $\partial T/ \partial t_i(a,k+1)$.
Suppose $D_{ij}(\cdot)$ and $C_i(\cdot)$ are known in closed-form, nodes can directly measure $D^\prime_{ij}(F_{ij})$ and $C^\prime_{i}(G_i)$ while transmitting on link $(i,j)$ and performing computation (or equivalently, first measure flows $F_{ij}$ and workloads $G_i$, then substitute into $D_{ij}(\cdot)$ and $C_i(\cdot)$). 
To collect $\partial T/ \partial t_i(a,k)$, we use recursive calculate \eqref{partial_D_r} starting with $i = d_a$ and $k = |\mathcal{T}_a|$ satisfying $\partial T / \partial t_{d_a}(a,|\mathcal{T}_a|) = 0$.
To carry out this recursive calculation, we apply a multi-stage distributed broadcast protocol to every application $a \in \mathcal{A}$, described as follows:

\begin{enumerate}
    \item Broadcast of $\partial T/\partial t_i(a,|\mathcal{T}_a|)$: 
    Each node $i$ first waits until it receives messages carrying $\partial T/\partial t_j(a,|\mathcal{T}_a|)$ from all its downstream neighbors $j\in\mathcal{N}(i)$ with $\phi_{ij}(a,|\mathcal{T}_a|) > 0$. 
    Then, node $i$ calculates $\partial T/\partial t_i(a,|\mathcal{T}_a|)$ by \eqref{partial_D_r_2} with the measured $D^\prime_{ij}(F_{ij})$ and received $\partial T/\partial t_j(a,|\mathcal{T}_a|)$. 
    Next, node $i$ broadcasts newly calculated $\partial T/\partial t_i(a,|\mathcal{T}_a|)$ to all its upstream neighbors $l\in \mathcal{N}(i)$ with $\phi_{li}(a,|\mathcal{T}_a|) > 0$. 
    (This stage starts with the destination $d_a$, where $d_a$ broadcasts $\partial T/\partial t_{d_a}(a,|\mathcal{T}_a|) = 0$ to its upstream neighbors.) 
     
    \item Broadcast of $\partial T/\partial t_i(a,k)$ for $k \neq |\mathcal{T}_a|$:
    (This stage starts with $k = |\mathcal{T}_a|-1$ and is repeated with $k$ being abstracted by $1$ each time, until $k = 0$.)
    Suppose every node $i$ has calculated $\partial T/\partial t_i(a,k^\prime)$ for all $k^\prime \geq k+1$. 
    Then, similar as stage 1), $\partial T/\partial t_i(a,k)$ can be calculated recursively by \eqref{partial_D_r_1} via broadcast.
    Besides $\partial T/\partial t_j(a,k)$ from all downstream neighbors $j$, node $i$ must also obtain $\partial T/\partial t_i(a,k+1)$ and $C^\prime_i(G_i)$ for \eqref{partial_D_r_1}.
    For each $k$, this stage starts at nodes $i$ where $\phi_{ij}(a,k) = 0$ for all $j \in \mathcal{V}$.{ i.e., the end-nodes of stage $(a,k)$ paths.} 
\end{enumerate}

With loop-free guaranteed, the broadcast procedure above is guaranteed to traverse throughout the network for all $(a,k) \in \mathcal{S}$ 
and terminate within a finite number of steps. 

\subsection{Blocked Nodes}
\label{subsection: Blocked nodes and scale matrices}
To achieve feasibility and the loop-free property, following \cite{gallager1977minimum}, 
we let $\mathcal{B}_i^t(a,k)$ be the set of nodes to which node $i$ is forbidden to forward any flow of stage $(a,k)$ at iteration $t+1$.

{
\begin{prop}
\label{prop_path}
Suppose $\boldsymbol{\phi}$ is a global optimal solution to \eqref{ServiceChainObj} satisfying \eqref{Condition_sufficient_chain}, then for any $(a,k) \in \mathcal{S}$ and $i \in \mathcal{V}$, if $\phi_{i0}(a,k) > 0$, then $\partial T/ \partial t_i(a,k) \geq \partial T / \partial t_{i}(a,k+1)$. 
Moreover, for all $j \in \mathcal{N}(i)$ with $\phi_{ij}(a,k) > 0$, it holds that $ \partial T / \partial t_i(a,k) \geq \partial T / \partial t_{j}(a,k)$, and this holds with strict inequality if $t_i(a,k) > 0$.
\end{prop}
\begin{proof}
    By Theorem \ref{Thm_Sufficient_chain}, suppose $\boldsymbol{\phi}$ satisfies condition \eqref{Condition_sufficient_chain}, then for any $i \in \mathcal{V}$, $(a,k) \in \mathcal{S}$ and $j \in \mathcal{N}(i) \cup \{0\}$, the following conditions are equivalent:
    \begin{enumerate}
        \item $\phi_{ij}(a,k) > 0$,
        \item $\delta_{ij}(a,k) \leq \delta_{ij^\prime}(a,k)$ for all $j^\prime \in \mathcal{N}(i) \cup \{0\}$.
    \end{enumerate}
   Therefore, sum \eqref{partial_D_r} over $j$ and recall \eqref{FlowConservation_phi}, it holds that 
   \begin{equation*}
       \partial T/ \partial t_{i}(a,k) = \delta_{ij}(a,k), \, \forall j\in \mathcal{N}(i) \cup \{0\}: \phi_{ij}(a,k)>0.
   \end{equation*}
   Recall that $D^\prime_{ij}(F_{ij}) \geq 0$ and $C^\prime_i(G_i) \geq 0$, thus for all $ j \in \mathcal{N}(i)$ with $\phi_{ij}(a,k)>0$,
   \begin{equation}
       \partial D/ \partial t_{i}(a,k) \geq \partial T/ \partial t_{j}(a,k),
       \label{prop_path_1}
   \end{equation}
   and if $\phi_{i0}(a,k) >0$, then
   \begin{equation}
       \partial T/\partial t_i(a,k) \geq \partial T /\partial t_{i}(a,k+1).
        \label{prop_path_2}
   \end{equation}
   Moreover, if $t_i(a,k)> 0$, it must hold $F_{ij} > 0$ if $\phi_{ij}(a,k)>0$, and $G_i > 0$ if $\phi_{i0}(a,k) > 0$. Combining with the monotonically increasing and convex property of $D_{ij}$ and $C_i$, we know $D^\prime_{ij}(F_{ij}) > 0$ and $C^\prime_i(G_i) > 0$, thus \eqref{prop_path_1} and \eqref{prop_path_2} hold with strict inequality. 
\end{proof}
}

The intuition for node blocking is as follows:
due to Proposition \ref{prop_path}, suppose $\boldsymbol{\phi}$ is a global optimal solution to \eqref{ServiceChainObj} satisfying \eqref{Condition_sufficient_chain}, then $\partial T/\partial t_i(a,k)$ should be monotonically decreasing along any path of stage $(a,k)$.
We thus require that during the algorithm, node $i$ should not forward any flow of stage $(a,k)$ to neighbor $j$ if either 1) $\partial T/\partial t_j(a,k) > \partial T/\partial t_i(a,k)$, or 2) it could form a path of stage $(a,k)$ containing some link $(p,q)$ such that $\partial T/\partial t_q(a,k) > \partial T/\partial t_p(a,k)$. 
The set containing nodes of these two categories, along with $j$ with $(i,j) \not\in \mathcal{E}$, is marked as the \emph{blocked node set} $\mathcal{B}_i^t(a,k)$.

Practically, the information needed to determine blocked node sets could be piggy-backed on the broadcast messages described in Section \ref{sec:broadcast} with light overhead.  
If the blocking mechanism is implemented for all $t$, the loop-free property is guaranteed to hold throughout the algorithm \cite{gallager1977minimum}. 


\subsection{Convergence and Complexity}
\label{sec:convergence}
The proposed algorithm can be implemented in an online fashion as it does not require prior knowledge of data input rates $r_i(a)$.
Moreover, it is adaptive to changes in $r_i(a)$ since flow rates $F_{ij}$ and workload $G_i$ can be directly estimated by the packet number on links and CPU in previous time slots. 
It can also adapt to changes in the network topology: whenever a link $(i,j)$ is removed from $\mathcal{E}$, node $i$ only needs to add $j$ to the blocked node set; when link $(i,j)$ is added to $\mathcal{E}$, node $i$ removes $j$ from the blocked node set.
{When a new node $v$ is added to the network, it can randomly initiate $\boldsymbol{\phi}_v$ with \eqref{FlowConservation_phi}. 
Then, $t_v(a,k)$ will be automatically updated by the marginal cost broadcast in the next time slot, and the loop-free property will be guaranteed.}

\begin{theo}
\label{thm_convergence}
Suppose the network starts at $\boldsymbol{\phi}^0$ with $T^0 < \infty$, and $\boldsymbol{\phi}^t$ is updated by Algorithm \ref{alg_GP} with a sufficiently small stepsize $\alpha$. 
Then, the sequence $\left\{\boldsymbol{\phi}^t\right\}_{t = 0}^{\infty}$ converges to a limit point $\boldsymbol{\phi}^*$ that satisfies condition \eqref{Condition_sufficient_chain}.
\end{theo}

We omit the proof of Theorem \ref{thm_convergence}, as it is a straightforward extension of \cite{gallager1977minimum} Theorem 5.
The stepsize $\alpha$ that guarantees convergence can also be found in \cite{gallager1977minimum}.
We remark that the convergence property of Algorithm \ref{alg_GP} can be improved by adopting second-order quasi-Newton methods, e.g., the method in \cite{xi2008node} and \cite{zhang2022optimal} speeds up convergence while guaranteeing convergence from any initial point. 
{Furthermore, by adopting Algroithm \ref{alg_GP} to adjust forwarding variables, the proof of Theorem \ref{thm:stability} is presented in Appendix \ref{Proof:Thm_stability}.}

Recall that $\boldsymbol{\phi}$ is updated every time slot of duration $T_{\text{slot}}$, and every broadcast message is sent once in every slot.
Thus there are $|\mathcal{E}|$ broadcast message transmissions for each stage in one slot, and totally $|\mathcal{S}||\mathcal{E}|$ per slot.
This implies there are on average $|\mathcal{S}|/T_{\text{slot}}$ broadcast messages per link/second, and at most $\Bar{d}|\mathcal{S}|$ for each node, where $\Bar{d}$ is the largest out-degree $\max_{v \in \mathcal{V}}|\mathcal{N}(v)|$.
We assume the broadcast messages are sent in an out-of-band channel.
Let $t_c$ be the maximum transmission time for a broadcast message, and
$\Bar{h}$ be the maximum hop number for all paths. 
The broadcast completion time for application $a$ is at most $(|\mathcal{T}_a|+1)\Bar{h}t_c$.
Moreover, the proposed algorithm has space complexity $O(|\mathcal{S}|)$ at each node. 

The update may fail if broadcast completion time $(|\mathcal{T}_a|+1)\Bar{h}t_c$ exceeds time slot length $T_{\text{slot}}$, or if the required broadcast bandwidth $|\mathcal{S}|/T_{\text{slot}}$ exceeds the broadcast channel capacity. 
If so, we can use longer slots.
We can also allow some nodes to perform lagged updates, or update every multiple slots.
Meanwhile, if $|\mathcal{S}|$ is large, the algorithm overhead can be significantly reduced by applying our algorithm only to the top applications causing most network traffic, while using other heuristic methods for the less important ones.

\section{Numerical Evaluation}
\label{Section:simulation}


In this section, we evaluate the proposed algorithm \textbf{GP} via a flow-level simulator available at \cite{Zhang_Joint-Routing-and-Computation-2022_2022}. 
We implement several baselines and compare the performance of those against \texttt{GP} over different networks and parameter settings.

We summarize the simulation scenarios in Table \ref{tab_scenario}. 
Algorithms are evaluated in the following network topologies: 
\textbf{Connected-ER} is a connectivity-guaranteed Erdős-Rényi graph, generated by creating links uniformly at random with probability $p = 0.1$ on a linear network concatenating all nodes.
\textbf{Balanced-tree} is a complete binary tree.
\textbf{Fog} is a sample topology for fog-computing, where nodes on the same layer are linearly linked in a balance tree \cite{kamran2019deco}.
\textbf{Abilene} is the topology of the predecessor of \emph{Internet2 Network} \cite{rossi2011caching}.
\textbf{LHC} is the topology for data sharing of the Large Hadron Collider at the European Organization for Nuclear Research (CERN).
\textbf{GEANT} is a pan-European data network for the research and education community \cite{rossi2011caching}.
\textbf{SW} (small-world) is a ring-like graph with additional short-range and long-range edges \cite{kleinberg2000small}.

Table \ref{tab_scenario} also summarizes the number of nodes $|\mathcal{V}|$, edges $|\mathcal{E}|$, and stages $|\mathcal{S}|$.
We assume each application has $R$ random active data sources (i.e., the nodes $i$ for which $r_i(a) > 0$), and $r_{i}(a)$ for each data source is chosen u.a.r. in $[0.5, 1.5]$.
\textbf{Link} is the type of communication costs $D_{ij}(\cdot)$, where \emph{Linear} denotes a linear cost $D_{ij}(F_{ij}) = d_{ij}F_{ij}$, and \emph{Queue} denotes a non-linear cost that represents the link queueing delay $D_{ij}(F_{ij}) = \frac{F_{ij}}{d_{ij} - F_{ij}}$.
\textbf{Comp} is the type of computation costs $C_i(G_i)$, where \emph{Linear} denotes the linear cost  $C_i(G_i) = s_iG_i$, and \emph{Queue} denotes a non-linear cost that represents the CPU queueing delay $C_i(G_i) = \frac{G_i}{s_i - G_i}$.

\begin{table}[htbp]
\footnotesize
\begin{tabular}{|c|p{0.01\textwidth}|p{0.01\textwidth}|p{0.01\textwidth}|p{0.01\textwidth}|c|p{0.01\textwidth}|c|p{0.015\textwidth}|}
\hline
\textbf{Network}&\multicolumn{8}{c|}{\textbf{Parameters}} \\
\textbf{Topology} & \multicolumn{1}{p{0.01\textwidth}}{$|\mathcal{V}|$} & \multicolumn{1}{p{0.01\textwidth}}{$|\mathcal{E}|$} & \multicolumn{1}{p{0.01\textwidth}}{$|\mathcal{A}|$} & \multicolumn{1}{p{0.01\textwidth}}{$R$} & \multicolumn{1}{c}{\textbf{Link}} & \multicolumn{1}{p{0.015\textwidth}}{$\Bar{d}_{ij}$} &\multicolumn{1}{c}{\textbf{Comp}}& $\Bar{s}_i$ \\
\hline
Connected-ER& \multicolumn{1}{p{0.01\textwidth}}{$20$} & \multicolumn{1}{p{0.01\textwidth}}{$40$} & \multicolumn{1}{p{0.01\textwidth}}{$5$} & \multicolumn{1}{p{0.01\textwidth}}{$3$} & \multicolumn{1}{c}{Queue} & \multicolumn{1}{p{0.015\textwidth}}{$10$} &\multicolumn{1}{c}{Queue}& $12$ \\
Balanced-tree & \multicolumn{1}{p{0.01\textwidth}}{$15$} & \multicolumn{1}{p{0.01\textwidth}}{$14$} & \multicolumn{1}{p{0.01\textwidth}}{$5$} & \multicolumn{1}{p{0.01\textwidth}}{$3$} & \multicolumn{1}{c}{Queue} & \multicolumn{1}{p{0.015\textwidth}}{$20$} &\multicolumn{1}{c}{Queue}& $15$ \\
Fog & \multicolumn{1}{p{0.01\textwidth}}{$19$} & \multicolumn{1}{p{0.01\textwidth}}{$30$} & \multicolumn{1}{p{0.01\textwidth}}{$5$} & \multicolumn{1}{p{0.01\textwidth}}{$3$} & \multicolumn{1}{c}{Queue} & \multicolumn{1}{p{0.015\textwidth}}{$20$} &\multicolumn{1}{c}{Queue}& $17$ \\
Abilene & \multicolumn{1}{p{0.01\textwidth}}{$11$} & \multicolumn{1}{p{0.01\textwidth}}{$14$} & \multicolumn{1}{p{0.01\textwidth}}{$3$} & \multicolumn{1}{p{0.01\textwidth}}{$3$} & \multicolumn{1}{c}{Queue} & \multicolumn{1}{p{0.015\textwidth}}{$15$} &\multicolumn{1}{c}{Queue}& $10$ \\
LHC & \multicolumn{1}{p{0.01\textwidth}}{$16$} & \multicolumn{1}{p{0.01\textwidth}}{$31$} & \multicolumn{1}{p{0.01\textwidth}}{$8$} & \multicolumn{1}{p{0.01\textwidth}}{$3$} & \multicolumn{1}{c}{Queue} & \multicolumn{1}{p{0.015\textwidth}}{$15$} &\multicolumn{1}{c}{Queue}& $15$ \\
GEANT & \multicolumn{1}{p{0.01\textwidth}}{$22$} & \multicolumn{1}{p{0.01\textwidth}}{$33$} & \multicolumn{1}{p{0.01\textwidth}}{$10$} & \multicolumn{1}{p{0.01\textwidth}}{$5$} & \multicolumn{1}{c}{Queue} & \multicolumn{1}{p{0.015\textwidth}}{$20$} &\multicolumn{1}{c}{Queue}& $20$ \\
SW & \multicolumn{1}{p{0.01\textwidth}}{$100$} & \multicolumn{1}{p{0.01\textwidth}}{$320$} & \multicolumn{1}{p{0.01\textwidth}}{$30$} & \multicolumn{1}{p{0.01\textwidth}}{$8$} & \multicolumn{1}{c}{(both)} & \multicolumn{1}{p{0.015\textwidth}}{$20$} &\multicolumn{1}{c}{(both)}& $20$ \\

\hline
\textbf{Other Parameters}&\multicolumn{8}{|c|}{
$|\mathcal{T}_a| = 2$, $r_i(a) \in [0.5,1.5]$, $L_{(a,k)} = 10 - 5k$
} \\

\hline
\end{tabular}
\vspace{-0.2\baselineskip}
\caption{Simulated Network Scenarios}
\label{tab_scenario}
\vspace{-0.5\baselineskip}
\end{table}

\begin{figure*}
\centerline{\includegraphics[width=0.95\textwidth]{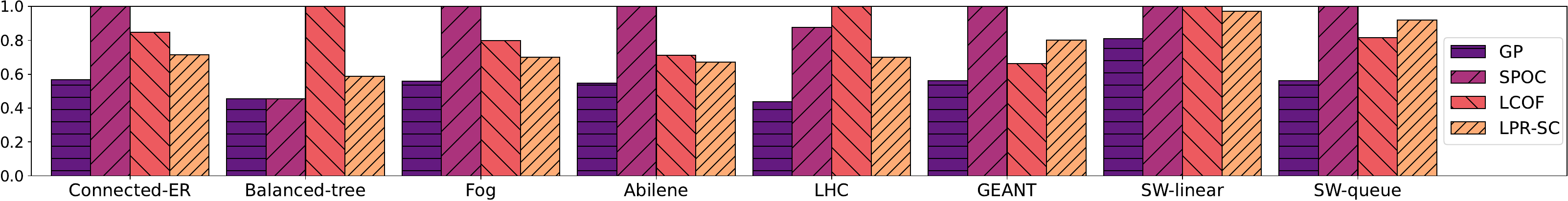}}
\caption{Normalized total cost for network scenarios in Table \ref{tab_scenario}}
\label{fig_bar}
\end{figure*}

\begin{figure}[h]
\begin{minipage}{0.49\linewidth}
\includegraphics[width=1\linewidth]{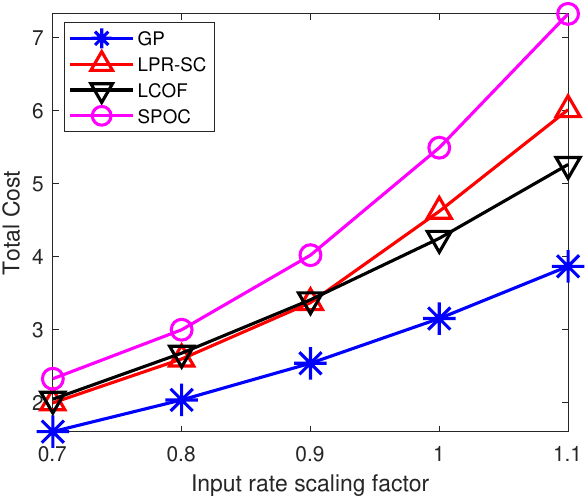}
\caption{$T(\boldsymbol{\phi})$ vs $r_i(a)$}
\label{fig:inputRate}
\end{minipage}
\begin{minipage}{0.49\linewidth}
\includegraphics[width=1\linewidth]{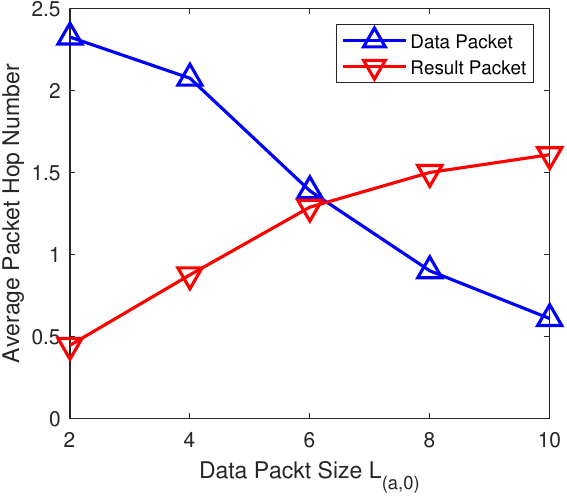}
\caption{Packet distance vs $L_{(a,0)}$ }
\label{fig:comploc}
\end{minipage}
\end{figure}

We implement the following baseline algorithms: 

\noindent\textbf{SPOC} (Shortest Path Optimal Computation placement) 
fixes the routing variables 
to the shortest path (measured with the marginal cost at $F_{ij} = 0$, accounting for the propagation delay without queueing effect), and studies the optimal computation placement along these paths. 
Namely, \texttt{SPOC} only optimize $T$ over variables $\phi_{i0}(a,k) \in [0,1]$. It sets $\phi_{ij} = 1-\phi_{i0}$ for $(i,j)$ on the shortest path, and sets $\phi_{ij} = 0$ for $(i,j)$ not on the shortest path.
A similar strategy is considered in \cite{he2021multi} with linear topology and computation partial offloading. 

\noindent\textbf{LCOF} (Local Computation Optimal Forwarding) computes all exogenous input flows at their data sources, 
 and optimally routes the result to destinations using \texttt{GP}.
That is, \texttt{LCOF} only optimizes $T$ over forwarding variables $\phi_{ij}(a,k)$ for $k = |\mathcal{T}_a|$ and $j \neq 0$. It sets all $\phi_{i0}(a,k) = 1$ at the data input sources for $k = 1,2,\cdots,|\mathcal{T}_a|-1$. 
Note that we focus on the scenarios where such pure-local computation is feasible, i.e., the computation costs are finite.  

\noindent\textbf{LPR-SC} (Linear Program Rounded for Service Chain) is the joint routing and offloading method by \cite{liu2020distributed}  (we extend this method heuristically to service chain applications), which does not consider partial offloading, congestible links and result flow. 
To adapt linear link costs in  \cite{liu2020distributed} to our schemes, we use the marginal cost at flow $F_{ij} = 0$. 

Fig.\ref{fig_bar} compares the total cost $T$ of different algorithms in steady state over networks in Table \ref{tab_scenario}, where the bar heights of each scenario are normalized according to the worst performing algorithm. 
We test both linear cost and nonlinear queueing delay with other parameters fixed in topology \texttt{SW}, labeled as \texttt{SW-linear} and \texttt{SW-queue}. 
The proposed algorithm \texttt{GP} significantly outperforms other baselines in all simulated scenarios, with as much as $50\%$ improvement over \texttt{LPR-SC}, which also jointly optimizes forwarding and computation offloading but does not consider congestible links. 
The difference of case \texttt{SW-linear} and \texttt{SW-queue} suggests that our proposed algorithm promises a more significant delay improvement for networks with nonlinear costs, e.g., when considering the queueing effect.
Note that \texttt{LCOF} and \texttt{SPOC} reflect the optimal objective for the packet forwarding and computation placement subproblems, respectively. The gain of jointly optimizing over both strategies could be inferred by comparing \texttt{GP} against \texttt{LCOF} and \texttt{SPOC}. For example, \texttt{LCOF} performs very poorly in topology \emph{Balanced-tree}, because the forwarding pattern cannot be optimized in a tree topology. 

We also perform refined experiments in \texttt{Abilene}.
Specifically, Fig.\ref{fig:inputRate} shows the change of total cost $T$ where all exogenous input rates $r_{i}(a)$ are scaled by the same factor, with other parameters fixed. 
The performance advantage of \texttt{GP} quickly grows as the network becomes more congested, especially against \texttt{LPR-SC}.

To further illustrate why \texttt{GP} outperforms baselines significantly with congestion-dependent cost, 
we define $H_{\text{data}}$ and $H_{\text{result}}$ as the average travel distance (hop number) of data packets from input to the first computation, and that of result packets from generation to being delivered, respectively.
In Fig. \ref{fig:comploc}, we compare $H_{\text{data}}$, $H_{\text{result}}$ for \texttt{GP} over the ratio of packet size (where the result size is fixed to $1$).
The trajectories suggest that the average computation offloading distance drops with the size of data packets. 
i.e., \texttt{GP} tends to place computation tasks that generate larger results nearer to the destination. 
This phenomenon demonstrates the underlying optimality of our proposed method, reaching a ``balance'' among the cost for data forwarding, result forwarding and computation, and therefore optimizing the total cost. 

\section{Extension: Congestion Control and Fairness}
\label{Section:extension}
Thus far, our method optimally solves joint forwarding and computation offloading problem \eqref{ServiceChainObj} for service chain applications, provided that the exogenous input request rates $\boldsymbol{r}$ lies in the stability region $\mathcal{D}_{\boldsymbol{r}}$.
There are practical situations, however, where the resulting network cost is excessive for given user demands even with the optimal forwarding and offloading strategy, since $\boldsymbol{r}$  may exceed the maximum network capacity. 
Moreover, the network operator may wish to actively balance the admitted rate from different users or applications, in order to achieve inter-user or inter-application fairness.
Therefore, to limit and balance the exogenous input rates, we extend our proposed framework by considering an extended graph to seamlessly incorporate a utility-based congestion control.
Our congestion control method is inspired by the idea in \cite{xi2008node} to accommodate service chain computation applications. 

\subsection{Utility-based fairness}
Extending the model in Section \ref{Section:model_chain} where the exogenous input rates $r_i(a)$ are pre-defined, in this section, we assume the network operator can actively control $r_i(a)$ within an interval $r_i(a) \in [0,\Bar{r}_i(a)]$ for all $i \in \mathcal{V}$ and $a \in \mathcal{A}$, where $\Bar{r}_i(a) \geq 0$ is a pre-defined constant upper limit specified by network users.

To mathematically capture the utility-based congestion control and fairness, we associate a \emph{utility function} $U_{ia}(\cdot)$ to the exogenous inputs, where the utility of user $i$'s input for application $a$ is given by $U_{ia}(r_i(a))$.

We assume the utility functions $U_{ia}(\cdot)$ are monotonically increasing and concave on $[0,\Bar{r}_i(a)]$ with $U_{ia}(0) = 0$.
Concave $U_{ia}(\cdot)$ subsumes a variety of commonly accepted utility and fairness metrics, and is widely adopted in the literature, e.g., \cite{liu2021fair}.
For example, the $\alpha$-fairness $U(r)$ parameterized by $\alpha \geq 0$ is given by
\begin{equation*}
    U(r) = \begin{cases}
        \frac{r^{1-\alpha}}{1-\alpha}, &\quad \text{ if } 0 \leq \alpha < 1
        \\ \log (r + \epsilon), &\quad \text{ if } \alpha = 1
        \\ \frac{(r + \epsilon)^{1-\alpha}}{1-\alpha}, &\quad \text{ if } \alpha > 1
    \end{cases}
\end{equation*}
where $\epsilon$ is a positive constant.
For any $\alpha \geq 0$, the $\alpha$-fairness $U(r)$ is concave in $r$, and is strictly concave if $\alpha >0$.

Incorporating the utility metrics $U_{ia}(\cdot)$, we seek to maximize a \emph{utility-minus-cost} following \cite{kelly1997charging}, defined as
\begin{equation}
\begin{aligned}
    &\max_{\boldsymbol{r},\boldsymbol{\phi}} \quad T(\boldsymbol{r},\boldsymbol{\phi}) = \sum_{i \in \mathcal{V}}\sum_{a\in\mathcal{A}} U_{ia}(r_i(a)) 
    \\ & \quad \quad \quad \quad - \sum_{(i,j)\in \mathcal{E}} D_{ij}(F_{ij}) - \sum_{i \in \mathcal{V}} C_{i}(G_i) 
    \\ &\text{subject to} \quad \boldsymbol{r} \in \mathcal{D}_{\boldsymbol{r}}, \quad  \boldsymbol{\phi} \in \mathcal{D}_{\boldsymbol{\phi}}(\boldsymbol{r}).
\end{aligned}
\label{CongestionControlObj}
\end{equation}

{
We remark that by assuming individual utilities for every combination $(i,a)$, we consider the \emph{inter-user inter-application} fairness, where the admitted rate of each user node $i$ and each application $a$ is balanced by maximizing the aggregated utility.
Alternatively, one could consider solely the \emph{inter-user} fairness by imposing utility $U_i(\cdot)$ on the total admitted rate at $i$ given by $\sum_{a}r_i(a)$, or solely the \emph{inter-application} fairness by imposing utility $U_{a}(\cdot)$ on the total admitted rate of application $a$ given by $\sum_i r_i(a)$.
In this paper, for the sake of mathematical integrity, we only study the \emph{inter-user inter-task} fairness in \eqref{CongestionControlObj}.
}

\subsection{Extended network}
Problem \eqref{CongestionControlObj} can be optimally solved by extending our network 
 model in Section \ref{Section:model_chain}. 
Consider an \emph{extended network} denoted by graph $\mathcal{G}^E = (\mathcal{V}^E,\mathcal{E}^{E})$,
where $\mathcal{V}^E = \mathcal{V} \cup \mathcal{V}^V$ denotes the physical nodes $\mathcal{V}$ and a set of \emph{virtual nodes} $\mathcal{V}^V$, 
and $\mathcal{E}^E =  \mathcal{E} \cup \mathcal{E}^V$ denotes the links in original set $\mathcal{E}$ and in the set of \emph{virtual links} $\mathcal{E}^V$.

The virtual node set $\mathcal{V}^V$ consists of $|\mathcal{V}|$ nodes, each corresponding to one of the physical node, serving as a ``gateway'' of request admission. We denote by $v^V$ the virtual node corresponding to physical node $v$. 
Set $\mathcal{E}^V$ consists of the virtual links coming out of the virtual nodes. Specifically, we assume the exogenous input requests are migrated from physical nodes to their corresponding virtual nodes, and the input rate of application $a$ at virtual node $i^V$ is fixed to the upper limit $\Bar{r}_i(a)$.
Virtual node $i^V$ has a virtual out-link $(i^V, i)$ connecting to the corresponding physical node, on which the actual admitted flow of rate $r_i(a)$ is forwarded.
For the remaining rate $(\Bar{r}_i(a) - r_i(a))$ that is rejected to the physical network, we assume it is admitted by $i^V$, and directly converted to the final result of stage $(a, |\mathcal{T}_a|)$ and sent to the destination $d_a$ through a virtual link $(i^V,d_a)$.
We denote by $f^V_{i}(a)$ the flow on the virtual link $(i^V,d_a)$, i.e., $f^V_i(a) = \Bar{r}_i(a) - r_i(a)$. 

Let $\phi_{i^V i}(a)$ denote the fraction of admitted rate at virtual node $i^V$ for application $a$, and let $\phi_{i^V d_a}(a)$ denote the fraction of rejected rate. 
If $\Bar{r}_i(a) > 0$, it holds that $\phi_{i^V i}(a) = r_i(a) / \Bar{r}_i(a)$ and $\phi_{i^V d_a}(a) = f^V_i(a) / \Bar{r}_i(a)$.
Then the flow conservation in \eqref{FlowConservation_phi} is augmented with
\begin{equation}
    \phi_{i^V i}(a) + \phi_{i^V d_a}(a) = 1, \quad \forall a \in \mathcal{A}, i \in \mathcal{V}. 
    \label{flowconservation_phi_congestion}
\end{equation}

\begin{figure}[t!]
\centerline{
\includegraphics[width=0.45\textwidth]{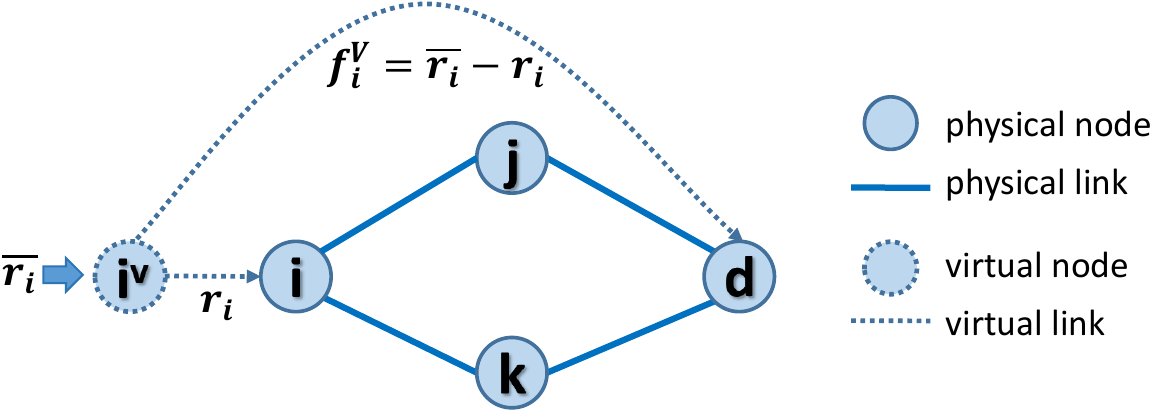}
}
\caption{Illustration of extended graph. The input rate upper limit $\Bar{r}_i$ is admitted by the virtual gate $i^{\text{V}}$. Among this, rate $r_i$ is further admitted by the physical node $i$. The rest is forwarded along the virtual link $(i^\text{V},d)$ and converted to the result stage.} 
\label{fig_congestion_control}
\end{figure}

We next assign link costs on the virtual links in $\mathcal{E}^V$.
For virtual link $(i^V,i)$, we do not assume any cost, i.e., $D_{i^Vi}(\cdot) \equiv 0$.
For virtual link $(i^V,d_a)$, we assume 
\begin{equation*}
\begin{aligned}
    D_{(i^V,d_a)}(f^V_i(a)) &= U_{ia}(\Bar{r}_i(a)) - U_{ia}(r_i(a))
    \\ &= U_{ia}(\Bar{r}_i(a)) - U_{ia}\left(\Bar{r}_i(a) - f^V_i(a)\right).
\end{aligned}
\end{equation*}
Namely, $D_{(i^V,d_a)}(f^V_i(a))$ represents the loss of utility due to congestion control.
Due to the concavity of $U_{ia}$, we know $D_{(i^V,d_a)}(f^V_i(a))$ is increasing convex in $f^V_i(a)$, coherent with our previous assumption on $D_{ij}(F_{ij})$ in Section \ref{Section:model_chain}.

Therefore, the utility-minus-cost maximization problem \eqref{CongestionControlObj} is equivalent to the following cost minimization problem on the extended graph $\mathcal{G}^E$, 
\begin{equation}
\begin{aligned}
        &\min_{\boldsymbol{\phi}^E} \quad T^E(\boldsymbol{\phi}^E) \equiv \sum_{(i,j)\in \mathcal{E}^E} D_{ij}(F_{ij}) + \sum_{i \in \mathcal{V}} C_{i}(G_i) 
    \\ &\text{subject to} \quad  \boldsymbol{\phi}^E \in \mathcal{D}_{\boldsymbol{\phi}^E},
\end{aligned}
\label{CongestionControlObj_eqv}
\end{equation}
where $\boldsymbol{\phi}^E$ includes both the physical forwarding variables $\boldsymbol{\phi}$ and the virtual node forwarding fractions $[\phi_{i^V i}(a), \phi_{i^V d_a}(a)]_{i,a}$. Set $\mathcal{D}_{\boldsymbol{\phi}^E}$ is defined by \eqref{FlowConservation_phi} and \eqref{flowconservation_phi_congestion}.

\subsection{Optimal congestion control}

Recall that condition \eqref{Condition_sufficient_chain} optimally solves \eqref{ServiceChainObj}, and observe that problem \eqref{CongestionControlObj_eqv} shares identical mathematical form with \eqref{ServiceChainObj}.
Therefore, we extend condition \eqref{Condition_sufficient_chain} to obtain an global optimal solution of \eqref{CongestionControlObj_eqv}.

\begin{theo}[Sufficient Condition with Congestion Control]
\label{Thm_Sufficient_congestion}
Let $\boldsymbol{\phi}^E \in \mathcal{D}_{\boldsymbol{\phi}^E}$.  
If the condition \eqref{Condition_sufficient_chain} holds for all $i \in \mathcal{V}$, $a \in \mathcal{A}$, $k = 0,\cdots,|\mathcal{T}_a|$, and the following holds for all $i^V \in \mathcal{V}^V$ and $a \in \mathcal{A}$, 
\begin{equation}
\begin{aligned}
    &\delta_{i^V i}(a) \leq \delta_{i^V d_a}(a), \quad \text{if } \phi_{i^V i}(a) > 0,
    \\ &\delta_{i^V i}(a) \geq \delta_{i^V d_a}(a), \quad \text{if } \phi_{i^V d_a}(a) > 0,
\end{aligned}
\label{Condition_sufficient_congestion}
\end{equation}
then $\boldsymbol{\phi}^E$ is a global optimal solution to \eqref{CongestionControlObj_eqv},
where
\begin{equation*}
\begin{aligned}
    \delta_{i^V i}(a) = \frac{\partial T}{ \partial t_i(a,0)}, \quad
     \delta_{i^V d_a}(a) = 
    U_{ia}^\prime(r_i(a)).
\end{aligned}
\end{equation*}
\end{theo}

The proof of Theorem \ref{Thm_Sufficient_congestion} is omitted as it is almost a repetition of Theorem \ref{Thm_Sufficient_chain}.
The above sufficient condition \eqref{Condition_sufficient_congestion} for congestion control can be intuitively interpreted as the following:
upon receiving a newly arrived input data packet application $a$ at node $i$, the congestion control gateway $i^V$ compares the marginal network cost if the packet is admitted, i.e., $\partial T/ \partial t_i(a,0)$, and the marginal utility loss if the packet is rejected, i.e., $U_{ia}^\prime(r_i(a))$.
The arrival packet is then admitted if the former is smaller and rejected if not.

The proposed Algorithm \ref{alg_GP} is naturally extendable to incorporate congestion control in a distributed and adaptive manner.
The implementation of each virtual node $i^V$ is carried out by the corresponding physical node $v$ with light overhead.
Remark that Algorithm \ref{alg_GP} itself does not specify how to find a feasible initial state $\boldsymbol{\phi}^0$. 
When extended to congestion control, however, a feasible initial state $(\boldsymbol{\phi}^E)^0$ is naturally introduced by setting $\phi_{i^V i}(a) = 0$ for all $i$ and $a$, i.e., the extended algorithm can always start with rejecting any arrival packets.

\section{Conclusion}
\label{Section:conclusion}
We propose a joint forwarding and computation offloading framework for service chain applications in edge computing networks with arbitrary topology and congestion-dependent nonlinear costs.
We formulate a non-convex total cost minimization problem, and optimally solve it by providing sufficient optimality conditions. 
We devise a distributed and adaptive online algorithm that reaches the global optimal.
Our method achieves delay-optimal forwarding and offloading for service chain computations, and can be applied to embed computation-intensive complex applications, e.g., DNN, into collaborative edge computing networks.
We also extend our method to consider network congestion control and utility-based fairness.
Our future work focuses on extending the proposed framework to incorporate general interdependency among tasks, e.g., directed acyclic graphs.


\begin{spacing}{0.9}
\bibliographystyle{IEEEtran}
\bibliography{References}
\end{spacing}

\appendix

\subsection{Proof of Lemma \ref{Lemma_Necessary_chain}}
\label{Proof:Lem_Necessary_chain}
The Lagrangian function of problem \eqref{ServiceChainObj} is given by
\begin{equation*}
    \begin{aligned}
        &L(\boldsymbol{\phi},\boldsymbol{\lambda},\boldsymbol{\mu}) = T(\boldsymbol{\phi}) - 
        \\ &\sum_{i \in \mathcal{V}}\sum_{ (a,k) \in \mathcal{S}} \lambda_{iak} \left( \sum_{j \in \{0\} \cup \mathcal{V}} \phi_{ij}(a,k) - \mathbbm{1}_{k \neq |\mathcal{T}_a| \text{ or } i \neq d_a} \right)
        \\ & - \sum_{i\in\mathcal{V}}\sum_{j\in\{0\}\cup\mathcal{V}}\sum_{ (a,k) \in \mathcal{S}} \mu_{ijak}\phi_{ij}(a,k),
    \end{aligned}
\end{equation*}
where $\boldsymbol{\lambda} = [\lambda_{iak}]$ and $\boldsymbol{\mu} = [\mu_{ijak}]$ with $\lambda_{iak} \in \mathbb{R}$ and $\mu_{ijak} \in \mathbb{R}^*$ are the Lagrangian multipliers corresponding to constraint \eqref{FlowConservation_phi} and $\phi_{ij}(a,k) \geq 0$, respectively.

Suppose $\boldsymbol{\phi}$ is a global optimal solution to \eqref{ServiceChainObj}, then there must exist a set of $(\boldsymbol{\lambda},\boldsymbol{\mu})$ such that \cite{bertsekas1997nonlinear}
\begin{equation*}
    \frac{\partial L}{\partial \phi_{ij}(a,k)} = 0, \quad \frac{\partial L}{\partial \lambda_{iak} = 0},\, \text{and } \mu_{ijak}\phi_{ij}(a,k) = 0.
\end{equation*}
By Section \ref{sec:optimality_chain}, for this set of $(\boldsymbol{\lambda},\boldsymbol{\mu})$, it holds that
\begin{equation*}
\begin{aligned}
   \frac{\partial L}{\partial \lambda_{iak}} &= \sum_{j \in \{0\} \cup \mathcal{V}} \phi_{ij}(a,k) - \mathbbm{1}_{k \neq |\mathcal{T}_a| \text{ or } i \neq d_a},
   \\ \frac{\partial L}{\partial \phi_{ij}(a,k)} &= \frac{\partial T}{\partial \phi_{ij}(a,k)} - \lambda_{iak} - \mu_{ijak}.
\end{aligned}
\end{equation*}
Combining above with the complementary slackness $\mu_{ijak}\phi_{ij}(a,k) = 0$ (i.e., when $\phi_{ij}(a,k) > 0$, it must hold that $\mu_{ijak} = 0$), and notice the arbitrariness of $\mu_{ijak}$ when $\phi_{ij}(a,k) = 0$, we know that 
\begin{equation*}
    \frac{\partial T}{\partial \phi_{ij}(a,k)} \begin{cases}
        = \lambda_{iak}, \quad \text{if } \phi_{ij}(a,k) > 0,
        \\> \lambda_{iak} \quad \text{if } \phi_{ij}(a,k) = 0.
    \end{cases}
\end{equation*}
Therefore, Lemma \ref{Lemma_Necessary_chain} holds by taking $\lambda_{iak} = \min_{j \in \{0\}\cup\mathcal{V}} \partial T/\partial \phi_{ij}(a,k)$ in the above.

\subsection{Proof of Theorem \ref{Thm_Sufficient_chain}.}
\label{Proof:Thm_sufficient_chain}
We follow the idea of \cite{gallager1977minimum} and \cite{zhang2022optimal} and generalize to service chain computation applications.
For simplicity, we focus throughout the proof on the situation when $\sum_{j}\phi_{ij}(a,k) = 1$ for every $i$ and $(a,k)$.
The final-stage-destination-node cases with $\sum_{j}\phi_{ij}(a,|\mathcal{T}_a|) = 0$ are omitted since all forwarding fractions must be $0$.

We assume strategy $\boldsymbol{\phi}$ satisfies condition \eqref{Condition_sufficient_chain}. To prove $\boldsymbol{\phi}$ optimally solves \eqref{ServiceChainObj}, let $\boldsymbol{\phi}^\dagger$ be any other feasible strategy such that $\boldsymbol{\phi} \neq \boldsymbol{\phi}^\dagger$, then it is sufficient to show that $T(\boldsymbol{\phi}^\dagger) \geq T(\boldsymbol{\phi})$.
We denote by $f_{ij}^\dagger(a,k)$ and $F_{ij}^\dagger$ the link flows under strategy $\boldsymbol{\phi}^\dagger$, as well as by $g_{i}^\dagger(a,k)$ and $G_i^\dagger$ the computation workloads.

Consider the following flow-based problem, which is equivalent to problem \eqref{ServiceChainObj}:
\begin{align}
    \min_{\boldsymbol{f}} \quad &T_{\boldsymbol{f}}(\boldsymbol{f}) = \sum_{(i,j) \in \mathcal{E}} D_{ij}(F_{ij}) + \sum_{i \in \mathcal{V}} C_i(G_i) \label{proof:flow_domain_problem}
    \\\text{s.t.} \quad & \sum_{j \in \mathcal{V}} f_{ji}(a,k) + g_i(a,k-1) = \sum_{j \in \mathcal{V}} f_{ij}(a,k) + g_i(a,k), \nonumber
    \\ & f_{ij}(a,k) \geq 0, \quad g_i(a,k) \geq 0, \nonumber
\end{align}
where $\boldsymbol{f} = \left[ f_{ij}(a,k),  g_i(a,k)\right]_{(a,k) \in \mathcal{S}, i,j \in \mathcal{V}}$ denotes the flow domain variable.
Note that $T_{\boldsymbol{f}}(\boldsymbol{f})$ is convex in $\boldsymbol{f}$, and the constraints of \eqref{proof:flow_domain_problem} form a convex polytope of $\boldsymbol{f}$. 
Therefore, let $\boldsymbol{f}^\dagger = \left[ f_{ij}^\dagger(a,k),  g_i^\dagger(a,k)\right]$, we know that for any $\alpha \in [0,1]$, $\boldsymbol{f}(\alpha) \equiv (1 - \alpha) \boldsymbol{f} + \alpha \boldsymbol{f}^\dagger$ is also feasible to \eqref{proof:flow_domain_problem}. 
We then let
\begin{equation*}
    \begin{aligned}
        T_\alpha(\alpha) = T_{\boldsymbol{f}}\left(\boldsymbol{f}(\alpha)\right).
    \end{aligned}
\end{equation*}

Then function $T_\alpha(\alpha)$ is also convex in $\alpha \in [0,1]$.
Recall that $T_\alpha(0) = T(\boldsymbol{\phi})$ and $T_\alpha(1) = T(\boldsymbol{\phi}^\dagger)$. To show $T(\boldsymbol{\phi}^\dagger) \geq T(\boldsymbol{\phi})$, it is sufficient to show 
\begin{equation}
    \frac{d T_{\alpha}(\alpha)}{d \alpha} \Big|_{\alpha = 0} \geq 0,
\label{proof:SuffCond_obj}
\end{equation}
where it is easy to see the LHS of \eqref{proof:SuffCond_obj} can be written as
\begin{equation}
\begin{aligned}
    \frac{d T_{\alpha}(\alpha)}{d \alpha} \Big|_{\alpha = 0} = &\sum_{(i,j) \in \mathcal{E}}D_{ij}^\prime(F_{ij})\left(F_{ij}^\dagger - F_{ij}\right) 
    \\ &+ \sum_{i \in \mathcal{V}}C_i^\prime(G_i)\left(G_i^\dagger - G_i\right).
\end{aligned}
\label{proof:SuffCond_derivative}
\end{equation}

For all $i \in \mathcal{V}$, $(a,k) \in \mathcal{S}$, let
\begin{equation*}
    \delta_i(a,k) = \min_{j^\prime \in \left\{0\right\}\cup\mathcal{V}} \delta_{ij}(a,k)
\end{equation*}
denote the minimal augmented marginal among all out-going directions, then by \eqref{Condition_sufficient_chain}, we have
\begin{equation*}
\begin{aligned}
        \frac{\partial T}{\partial t_i(a,k)} &= \sum_{j \in \left\{0\right\}\cup\mathcal{V}} \phi_{ij}(a,k) \delta_{ij}(a,k)
        \\ &= \sum_{j \in \left\{0\right\}\cup\mathcal{V} : \phi_{ij}(a,k) > 0}  \phi_{ij}(a,k)\delta_i(a,k)
        \\ &= \delta_i(a,k)
\end{aligned}
\end{equation*}
and thus for all $i \in \mathcal{V}$, $(a,k) \in \mathcal{S}$ and $j \in \left\{0\right\}\cup\mathcal{V}$,
\begin{equation}
    \delta_{ij}(a,k) \geq \frac{\partial T}{\partial t_i(a,k)}.
\label{proof:SuffCond1}
\end{equation}

Multiply both side of \eqref{proof:SuffCond1} by $\phi_{ij}^\dagger(a,k)$ and sum over $j \in \left\{0\right\}\cup\mathcal{V}$, combining with \eqref{delta}, we have
\begin{equation}
    \begin{aligned}
        \sum_{j \in \mathcal{V}} & L_{(a,k)} D^\prime_{ij}(F_{ij})\phi_{ij}^\dagger(a,k) + w_i(a,k) C^\prime_i(G_i)\phi_{i0}^\dagger(a,k)
        \\ &\geq \frac{\partial T}{\partial t_i(a,k)} - \sum_{j \in \mathcal{V}} \frac{\partial T}{\partial t_j(a,k)}\phi_{ij}^\dagger(a,k) 
        \\ &- \frac{\partial T}{\partial t_i(a,k+1)} \phi_{i0}^\dagger(a,k),
    \end{aligned}
\label{proof:SuffCond2}
\end{equation}

Then multiply both side of \eqref{proof:SuffCond2} by $t_i^\dagger(a,k)$, recall that $f_{ij}(a,k) = t_i(a,k)\phi_{ij}(a,k)$ and $g_i(a,k) = t_i(a,k)\phi_{i0}(a,k)$, we get
\begin{equation}
    \begin{aligned}
        &\sum_{j \in \mathcal{V}} L_{(a,k)}D^\prime_{ij}(F_{ij})f_{ij}^\dagger(a,k) + w_i(a,k)C^\prime_i(G_i)g_{i}^\dagger(a,k)
        \\ &\geq t^\dagger_i(a,k) \frac{\partial T}{\partial t_i(a,k)} - \sum_{j \in \mathcal{V}} t^\dagger_i(a,k) \frac{\partial T}{\partial t_j(a,k)}\phi_{ij}^\dagger(a,k)
        \\ &- {t^\dagger_i(a,k)} \frac{\partial T}{\partial t_i(a,k+1)} \phi_{i0}^\dagger(a,k).
    \end{aligned}
    \label{proof:SuffCond3}
\end{equation}

Recall that for $k = \left|\mathcal{T}_a\right|$ there is no flow forwarded to the computation unit, we write $g_i(a,\left|\mathcal{T}_a\right|) \equiv 0$. 
Sum both sides of \eqref{proof:SuffCond3} over $(a,k) \in \mathcal{S}$, $i \in \mathcal{V}$ and rearrange, it holds that
\begin{equation}
    \begin{aligned}
        &\sum_{(i,j) \in \mathcal{E}} D_{ij}^\prime(F_{ij}) F_{ij}^\dagger + \sum_{i \in \mathcal{V}}C^\prime_i(G_i)G_i^{\dagger}
        \\ &\geq \sum_{i \in \mathcal{V}} \sum_{a \in \mathcal{A}} \sum_{k = 0}^{\left|\mathcal{T}_a\right|} t_i^\dagger(a,k)\frac{\partial T}{\partial t_i(a,k)}
        \\&- \sum_{i \in \mathcal{V}} \sum_{a \in \mathcal{A}} \sum_{k = 0}^{\left|\mathcal{T}_a\right|} {t^\dagger_i(a,k)} \frac{\partial T}{\partial t_i(a,k+1)} \phi_{i0}^\dagger(a,k)
        \\ &- \sum_{a \in \mathcal{A}} \sum_{k = 0}^{\left|\mathcal{T}_a\right|} \sum_{j \in \mathcal{V}} \frac{\partial T}{\partial t_j(a,k)} \left(\sum_{i \in \mathcal{N}(j)} t^\dagger_i(a,k) \phi_{ij}^\dagger(a,k)\right)
    \end{aligned}
    \label{proof:SuffCond4}
\end{equation}

Meanwhile, since the flow conservation
\begin{equation*}
    \sum_{j \in \mathcal{V}} f_{ji}(a,k) + g_i(a,k-1) = \sum_{j \in \mathcal{V}} f_{ij}(a,k) + g_i(a,k)
\end{equation*}
also applies for strategy $\boldsymbol{\phi}^\dagger$, we know for all $j \in \mathcal{V}$, $(a,k) \in \mathcal{S}$,
\begin{equation*}
    \sum_{i \in \mathcal{N}(j)} t_i^\dagger(a,k)\phi_{ij}^\dagger(a,k) = t_j^\dagger(a,k) - g_j^\dagger(a,k-1) - g_j^\dagger(a,k).
\end{equation*}

Substitute above into the right most term of \eqref{proof:SuffCond4}, we get
\begin{equation}
    \begin{aligned}
        &\sum_{(i,j) \in \mathcal{E}} D_{ij}^\prime(F_{ij}) F_{ij}^\dagger + \sum_{i \in \mathcal{V}}C^\prime_i(G_i)G_i^\dagger
        \\ &\geq \sum_{a \in \mathcal{A}} \sum_{k = 0}^{\left|\mathcal{T}_a\right|} \sum_{i \in \mathcal{V}} \frac{\partial T}{\partial t_i(a,k)}g_i^\dagger(a,k-1) 
        \\&+  \sum_{a \in \mathcal{A}} \sum_{k = 0}^{\left|\mathcal{T}_a\right|} \sum_{j \in \mathcal{V}} \frac{\partial T}{\partial t_i(a,k)}g_i^\dagger(a,k)
        \\ & -\sum_{i \in \mathcal{V}} \sum_{a \in \mathcal{A}} \sum_{k = 0}^{\left|\mathcal{T}_a\right|} {t^\dagger_i(a,k)}\frac{\partial T}{\partial t_i(a,k+1)} \phi_{i0}^\dagger(a,k).
    \end{aligned}
\label{proof:SuffCond5}
\end{equation}

We assumed for coherent that $g_i(a,k-1) \equiv r_i(a)$ if $k = 0$, and that $\phi_{i0}(a,\left|\mathcal{T}_a\right|) \equiv 0$, then \eqref{proof:SuffCond5} is equivalent to
\begin{equation}
    \begin{aligned}
        &\sum_{(i,j) \in \mathcal{E}} D_{ij}^\prime(F_{ij}) F_{ij}^\dagger + \sum_{i \in \mathcal{V}}C^\prime_i(G_i) G_i^\dagger
        \\ &\geq \sum_{a \in \mathcal{A}} \sum_{k = 0}^{\left|\mathcal{T}_a\right|} \sum_{i \in \mathcal{V}} \frac{\partial T}{\partial t_i(a,k)}g_i(a,k-1) 
        \\ &+ \sum_{i \in \mathcal{V}} \sum_{a \in \mathcal{A}} \Bigg(\sum_{k = 0}^{\left|\mathcal{T}_a\right|}\frac{\partial T}{\partial t_i(a,k)}g_i^\dagger(a,k-1) 
        \\&- \sum_{k = 0}^{\left|\mathcal{T}_a\right|}\frac{\partial T}{\partial t_i(a,k+1)}g_i^\dagger(a,k)\Bigg)
        \\ &= \sum_{i \in \mathcal{V}} \sum_{(a,k) \in \mathcal{S}} \frac{\partial T}{\partial t_i(a,k)}g_i(a,k-1).
    \end{aligned}
\label{proof:SuffCond6}
\end{equation}

On the other hand, \eqref{proof:SuffCond2} will hold with equality if substitute $\boldsymbol{\phi}^\dagger$ with $\boldsymbol{\phi}$, since \eqref{proof:SuffCond1} hold with equality for $\phi_{ij}(a,k) > 0$. Therefore, the derivation \eqref{proof:SuffCond3}\eqref{proof:SuffCond4}\eqref{proof:SuffCond5} and \eqref{proof:SuffCond6} will hold with equality for $\boldsymbol{\phi}$.
Specifically, substituting $\boldsymbol{\phi}^\dagger$ with $\boldsymbol{\phi}$, we have the following analogue of \eqref{proof:SuffCond6},
\begin{equation}
    \begin{aligned}
        &\sum_{(i,j) \in \mathcal{E}} D_{ij}^\prime(F_{ij}) F_{ij} + \sum_{i \in \mathcal{V}}C^\prime_i(G_i) G_i 
        \\&= \sum_{i \in \mathcal{V}} \sum_{(a,k) \in \mathcal{S}} \frac{\partial T}{\partial t_i(a,k)}g_i(a,k-1).
    \end{aligned}
\label{proof:SuffCond7}
\end{equation}

Subtracting \eqref{proof:SuffCond7} from \eqref{proof:SuffCond6}, we get
\begin{equation*}
     \sum_{(i,j) \in \mathcal{E}}D_{ij}^\prime(F_{ij})\left(F_{ij}^\dagger - F_{ij}\right) + \sum_{i \in \mathcal{V}}C_i^\prime(G_i)\left(G_i^\dagger - G_i\right) \geq 0,
\end{equation*}
which combined with \eqref{proof:SuffCond_derivative} complete the proof of \eqref{proof:SuffCond_obj}.


\subsection{Proof of Theorem \ref{thm:stability}}
\label{Proof:Thm_stability}
For analytical simplicity, we only prove for the case where $\Delta \boldsymbol{r}$ has only one non-zero element, that is, only $\Delta r_{i}(a,k) \neq 0$.
Without loss of generality, we assume $|\mathcal{A}| = 1$ and $|\mathcal{T}_a| = 0$ for that $a \in \mathcal{A}$, i.e., the network only performs packet forwarding for only one application, without the need to conduct any computation. Therefore, we omit the notation $(a,k)$ for simplicity.\footnote{For the case with computation and multiple applications, our analysis can be generalized as we essentially treat computation steps as special links.}
We further assume all communication cost functions $D_{ij}(\cdot)$ are strictly convex.

Suppose $\boldsymbol{\phi}$ satisfies the sufficient condition \eqref{Condition_sufficient_chain} with input rate $\boldsymbol{r}$, it is evident that $\boldsymbol{\phi}$ is loop-free.
When the input rate $r_i$ is increased by a sufficiently small $\Delta r_i$ ($\Delta r_i$ can be positive or negative, as long as the new input rate vector lies within stability region $\mathcal{D}_{\boldsymbol{r}}$. Without loss of generality, we assume $\Delta r_i > 0$), we apply Algorithm \ref{alg_GP} one node at a time, starting from node $i$.

Consider the change of $[\delta_{ij}]$ for all $j$ when the forwarding strategy $\boldsymbol{\phi}$ is kept unchanged.
To break down the problem, consider the DAG (directed acyclic graph) constructed by $(i,j)\in\mathcal{E}$ such that $\phi_{ij} > 0$.

(1) If node $i$ is the destination node $d$, then $\delta_{ij}$ is not changed on any link.

(2) If node $i$ is one-hop away from $d$, then it must hold that $\phi_{id} = 1$ and $\phi_{ij} = 0$ for all other $j$. In this case, $\delta_{id} = D_{id}^\prime(f_{id})$, and $$\Delta \delta_{id} = D_{id}^\prime(f_{id} + \Delta r_i) - D_{id}^\prime(f_{id}) = \Delta r_i D^{\prime\prime}_{id}(f_{id}).$$
Therefore, we let $\delta^\prime_{id}$ be the marginal increase of $\delta_{id}$ due to the increase of $r_i$, and $$\delta^\prime_{id} = \frac{\Delta \delta_{id}}{\Delta r_i} = D^{\prime\prime}_{id}(f_{id}).$$
Moreover, let $\delta_i = \sum_{j} \phi_{ij} \delta_{ij}$, then $\delta^\prime_i = \sum_{j} \phi_{ij} \delta^\prime_{ij}$. In this case, $\delta_i^\prime = D^{\prime\prime}_{id}(f_{id})$.

(3) If node $i$ is more than one-hop away from $d$, then it holds that
\begin{equation}
\begin{aligned}
    &\delta_{ij}^\prime = D^{\prime\prime}_{ij}(f_{ij}) + \delta^\prime_j,
    \\& \delta_i^\prime =  \sum_{j} \phi_{ij} \delta^\prime_{ij}.
\end{aligned}
\label{proof_second_deri}
\end{equation}
We denote all paths from node $i$ to $d$ in the DAG by set $\mathcal{P}_i$, where each path $p \in \mathcal{P}_i$ is a sequence of nodes $(p_1,p_2,\cdots,p_{|p|})$ with $p_1 = i$, $p_{|p|} = d$ and $\phi_{p_k p_{k+1}} >0$ for $k = 1,\cdots,|p|-1$.
Therefore, by recursively applying \eqref{proof_second_deri} from node $d$ to node $i$ on the reverse direction for all path in $\mathcal{P}_i$, we have
\begin{equation}
\begin{aligned}
    \delta^\prime_i &= \sum_{p \in \mathcal{P}_i} \left(D^{\prime\prime}_{p_1p_2} + \left(D^{\prime\prime}_{p_2p_3} + \cdots \right)\phi_{p_2p_3}\right)\phi_{p_1p_2}
    \\ &= \sum_{p \in \mathcal{P}_i} \left(\sum_{k = 1}^{|p|-1} D^{\prime\prime}_{p_kp_{k+1}}\prod_{l = 1}^k \phi_{p_lp_{l+1}}\right)
\end{aligned}
\label{proof_delta_prime}
\end{equation}

Combining the above cases, we know that for any $i$, when $r_i$ in increased by a small amount $\Delta r_i$, the marginal cost $\delta_{ij}$ for all $j$ is increased by
\begin{equation*}
    \Delta \delta_{ij} = \Delta r_i \left(D^{\prime\prime}_{ij} + \sum_{p \in \mathcal{P}_j} \left(\sum_{k = 1}^{|p|-1} D^{\prime\prime}_{p_kp_{k+1}}\prod_{l = 1}^k \phi_{p_lp_{l+1}}\right) \right).
\end{equation*}
Therefore, recall the algorithm update \eqref{dphi_and_dy}\eqref{algorithm_detail_eNS}, and combined with the fact that $\boldsymbol{\phi}$ already satisfies \eqref{Condition_sufficient_chain}, 
the adjust amount $\Delta \phi_{ij} \equiv \phi^1_{ij} - \phi_{ij}$ is upper bounded by 
\begin{equation}
\label{proof_bound}
    \Delta \phi_{ij} \leq \alpha \Delta \delta_{ij}. 
\end{equation}
Moreover, it is shown by \cite{gallager1977minimum} that Algorithm \ref{alg_GP} converges linearly to the optimal solution satisfying \eqref{Condition_sufficient_chain}.
By adopting section order methods, e.g., \cite{xi2008node,bertsekas1984second}, the rate of convergence can be enhanced to super-linear.
Therefore, let $\boldsymbol{\phi}^*$ be the convergent solution of Algorithm \ref{alg_GP} after introducing $\Delta r_i$, there exist a finite scalar $M$ that
\begin{equation}
\label{proof_distance_bound}
    {|\boldsymbol{\phi}^{t+1} - \boldsymbol{\phi}^*|} \leq M {|\boldsymbol{\phi}^{t} - \boldsymbol{\phi}^*|},
\end{equation}
and there exists a scalar $\mu < 1$ such that 
\begin{equation}
\label{proof_convergence_rate}
    \lim_{t \to \infty} \frac{|\boldsymbol{\phi}^{t+1} - \boldsymbol{\phi}^*|}{|\boldsymbol{\phi}^{t} - \boldsymbol{\phi}^*|} = \mu.
\end{equation}

Combining \eqref{proof_bound}\eqref{proof_distance_bound}\eqref{proof_convergence_rate}, there exists a finite constant $C$ such that $| \boldsymbol{\phi}^* - \boldsymbol{\phi}| \leq C\Delta r_i$, i.e., $$\lim_{\Delta r_i \to 0}| \boldsymbol{\phi}^* - \boldsymbol{\phi}| = 0.$$

Therefore, combing back to Theorem \ref{thm:stability}, we know there exist a function $\epsilon(\delta)$ for $\delta > 0$ that is continuous at $0$, 
and for $\Delta r_i$ that $\left|\Delta r_i\right| < \delta$, the corresponding new optimal solution $\boldsymbol{\phi}^*$ satisfies $| \boldsymbol{\phi}^* - \boldsymbol{\phi}| \leq \epsilon(\delta)$.
To generalize to multiple applications or multiple non-zero input rate $r_i$ changes, the analysis above still holds, as the constant $C$ would be the sum of all applications and all $\Delta r_i$, however, still finite.
To generalize to computation placement, one only needs to consider each computation step as a special link that goes back to the computation node itself, with a link cost associated.


\end{document}